%% file: sigconst.tex
\documentclass[aps,prd,showpacs,superscriptaddress,reprint]{revtex4-1}
\usepackage{amsmath,amsfonts,amssymb}
\usepackage{dsfont,graphicx}
\usepackage{bm}
\graphicspath{{Plots/}}
\newcommand{\bs}{\begin{subequations}}
\newcommand{\es}{\end{subequations}}
\newcommand{\adb}{\allowdisplaybreaks } 
\newcommand{\ann}{\adb \nonumber \\}
\newcommand{\TE}{\mathsf{TE}}
\newcommand{\TM}{\mathsf{TM}}
\newcommand{\num}{\textsf{num}}
\newcommand{\as}{\textsf{as}}
\newcommand{\ren}{\textsf{ren}}
\usepackage{color} \definecolor{darkgreen}{rgb}{0,.5,0}
\usepackage[colorlinks,filecolor=blue,citecolor=darkgreen,unicode]{hyperref}
\begin{document}
\title{Casimir energy for surfaces with constant conductivity}

\author{Nail Khusnutdinov}\email{nail.khusnutdinov@gmail.com}
\affiliation{ Institute of Physics, Kazan Federal University, Kremlevskaya 18, Kazan, 420008, Russia}
\affiliation{Department of Physics, University of South Florida, Tampa, Florida 33620, USA}
\author{D. Drosdoff}
\affiliation{Department of Physics, University of South Florida, Tampa, Florida 33620, USA}
\author{Lilia M. Woods}
\affiliation{Department of Physics, University of South Florida, Tampa, Florida 33620, USA}
\date{\today}
\begin{abstract}
We consider the vacuum energy of the electromagnetic field in systems characterized by a constant conductivity using the zeta-regularization approach. The interaction in two cases is investigated: two infinitely thin parallel sheets and an infinitely thin spherical shell. We found that the Casimir energy for the planar system is always attractive and it has the same characteristic distance dependence as the interaction for two perfect semi-infinite metals. The Casimir energy for the spherical shell  depends on the inverse radius of the sphere, but it maybe negative or positive depending on the value of the conductivity. If the conductivity is less than a certain critical value, the interaction is attractive, otherwise the Casimir force is repulsive regardless of the spherical shell radius.
\end{abstract}  
\maketitle 

\section{Introduction}

Since the seminal Casimir papers \cite{Casimir:1948:Main, *Casimir:1948:TIrLdWf}, great progress has been made in theoretical descriptions and experimental demonstrations of the Casimir effect.  Although this phenomenon has been studied extensively, its fundamental understanding still remains elusive due to the complicated relationship between the geometry, dielectric properties and temperature of the objects \cite{Bordag:2001:NdCe,Milton:2004:TCercap,Buhmann:2007:Dfimqe,Klimchitskaya:2009:TCfbrmet,Milton:2001:CEPMZE,Bordag:2009:ACE,Banishev:2013:DoACFbC}. This effect is a direct manifestation of the quantum fluctuations in vacuum. For smaller distances retardation from the finite speed of light is not important, thus the interaction becomes equivalent to the van der Waals force. 

Many investigations focus on objects with planar, spherical or cylindrical shapes \cite{Buhmann:2007:Dfimqe,Bordag:2009:ACE}. Recently an approach using functional determinants \cite{Emig:2006:CIbaPaaC,*Bulgac:2006:SCebDsops} has been proposed, which may be used for systems with more complicated extensions. In addition to the geometry, the response properties are also important for the Casimir force. Different models, including an infinite conductivity, and the plasma and Drude-Lorentz models for the dielectric function have been utilized, highlighting different functionalities of this phenomenon \cite{Bordag:2009:ACE}.

The recent advent of research in surface materials, such as graphene \cite{Novoselov:2004:EFEiATCF}, has been beneficial for studies of long ranged interactions in reduced dimensions. It has been shown that vdW/Casimir forces are of great importance to how graphene interacts with other systems \cite{Dobson:2006:ADIABvdWEF,*Gomez-Santos:2009:TvdWibgl,*Drosdoff:2010:Cfgs,*Drosdoff:2012:EosdotCfbgs,*Klimchitskaya:2013:vdWaCibtgs,*Banishev:2013:MtCfgfgoaSIOS,%
Bordag:2006:LffgascnvdWaCi,*Bordag:2009:CibapcagdbtDm,*Drosdoff:2011:Cibgsam,*Phan:2012:Ioagswafmp,*Bordag:2012:TCeitiogwdam,%
Dino:2004:DAatZEoG,*Bondarev:2004:vdWeusacdcn,%
*Bondarev:2005:vdWciadcn,*Blagov:2007:vdWibamaascn,*Churkin:2010:CohaDmodibgaHHoNa,*Chaichian:2012:TCiodawg}. 

Several reports have investigated graphene Casimir interaction in the context of a hydrodynamic or plasma models for the dielectric response using a dyadic Green's function technique or the standard Lifshitz approach \cite{Barton:2004:Cesps,*Barton:2005:CefpsIE,Bordag:2006:TCeftpsatrotsp,*Bordag:1992:Veiqftwepcop,Bordag:2007:Ioacwatps,%
Fetter:1973:EoalegISl,*Fetter:1974:EoalegIPa}.  The Dirac nature of the carriers and the unique optical properties have important implications for the graphene vdW/Casimir interaction. It has been shown that the graphene conductivity is a constant ($\sigma_{gr} = e^2/4\hbar$) over a relatively large frequency range, near infrared to optical     \cite{Gusynin:2007:Mcig,*Falkovsky:2007:Sdogc,*Nair:2008:FSCDVToG}. As a result, the graphene/graphene Casimir interaction has similar distance dependence as two perfect metals,  but the  magnitude is much reduced. At the same time, the interaction is dominated by thermal fluctuations at separations larger than 50 $nm$ \cite{Dobson:2006:ADIABvdWEF,*Gomez-Santos:2009:TvdWibgl,*Drosdoff:2010:Cfgs,*Drosdoff:2012:EosdotCfbgs,*Klimchitskaya:2013:vdWaCibtgs,*Banishev:2013:MtCfgfgoaSIOS}.

The zero-point photon energy summation method is another way to compute the Casimir energy and it is especially useful when the medium absorption is less important and can be neglected. This method involves a contour integration in the complex frequency plane of two sums, which are divergent. Removing the divergences requires making use of different regularization schemes either via the generalized $\zeta$-function or exponential cutoffs \cite{Bordag:2009:ACE}. The mode summation method is an elegant way to obtain Casimir interactions for objects with cylindrical and spherical shapes \cite{Bordag:2009:ACE} as well. Most of the studies have utilized either perfect conductors or dielectric-diamagnetic media \cite{Bordag:2009:ACE} to facilitate the calculations.  

The purpose of this investigation is to expand the use of the mode summation technique for systems with constant conductivity. In particular, we are interested in the consequences of planar and spherical shapes with $\sigma=const$ on the Casimir interaction. Dimensional analysis shows that for planar systems the energy due to this is a quantum-mechanical relativistic phenomenon that can be expressed as $\mathcal{E}^{p} =\hbar c Q^{p}(\bar \sigma)/d^3$, while for spherical systems --  $\mathcal{E}^{s} = \hbar c Q^{s}(\bar \sigma)/R$. 
$Q$ is a function only of $\bar\sigma = \sigma /c$. Our subsequent calculations confirm these expressions and show that while $\bar\sigma = \sigma /c$ is always positive, $Q^{s}(\bar \sigma)$ can change its sign as a function of $\bar \sigma$. Another surprising result is that all obtained expressions contain no singularities, which is in contrast to all other previously studied cases, where the renormalization of parameters of the classical part of the energy  has been used to obtain finite results. This is especially useful to widen the use of the zero point summation method since the main obstacle for its application has been the need for regularization techniques, which are not universal and system dependent. 

The paper is organized as follows. In Sec. \ref{Sec:PlSym} we derive the boundary conditions for the electromagnetic fields for two planar sheets with constant conductivity. The Casimir energy is calculated via the mode summation method. Section \ref{Sec:SpSym} is devoted to the electromagnetic boundary conditions and mode summation method energy calculation for a spherical shell with a constant conductivity. Discussion are given in Sec. \ref{Sec:Disc}. 

\section{Planar symmetry}\label{Sec:PlSym}

The first system under consideration consists of two parallel planar surfaces separated by a distance $d$ along the vertical direction. The two surfaces are inserted in a large box with boundaries $\pm L$ along the vertical with Dirichlet boundary conditions. Each surface is also characterized by a constant isotropic conductivity $\sigma=const$. The Casimir energy of this system can be expressed in terms of an infinite sum of zero-point photon energies. The energy per unit area has contributions from the transverse electric (TE) and transverse magnetic (TM) modes as follows:
\begin{equation}
\mathcal{E}_{\TM,\TE}^{p} = \frac{\hbar}{2} \iint \frac{d^2k_\perp}{(2\pi)^2}\sum_{j} (\omega_{j}^{\TM,\TE} -\omega_{j}^{\TM,\TE}|_{d\to \infty}), \label{eq:EPSDef}
\end{equation}  
where the integration is done over the planar wave vector. The second term in Eq. (\ref{eq:EPSDef}) corresponds to the reference vacuum when the planar layers are taken far apart. This is a divergent expression, which can be regularized in the framework of the zeta-regularization procedure represented as a limit of an analytical function
\begin{gather*}
\mathcal{E}_{\TM,\TE}^{p}=\lim_{s\to 0} \mathcal{E}_{\TM,\TE}^{p}(s)\ann
= \lim_{s\to 0} \frac{\hbar}{2} \Omega^{2s}\iint \frac{d^2k_\perp}{(2\pi)^2}\sum_{j} \left(\omega_{j}^{\TM,\TE}\right)^{1-2s}.
\end{gather*}
The parameter $\Omega$ with a  frequency dimension is introduced to keep the energy dimension of $\mathcal{E}(s)$. 

By converting the sum into a contour integral over imaginary frequency axis \cite{Bordag:1997:Cefmfitb,Bordag:2009:ACE}, the energy can further be transformed 
\begin{eqnarray}
\mathcal{E}_{\TM,\TE}^{p}(s) &=& - \hbar c \Lambda^{2s} \frac{\cos\pi s}{2\pi} \iint \frac{d^2k_\perp}{(2\pi)^2} \int_0^\infty d\lambda \lambda^{1-2s}\ann
&\times& \frac{\partial}{ \partial \lambda} \ln \Psi_{\TM,\TE}(i\lambda c), \label{eq:EtetmPlanar}
\end{eqnarray}
where $\Lambda = \Omega c$ and $\Psi_{\TM,\TE}$ are the photon energy spectra for the TE and TM modes. 

As was noted in Ref. \cite{Barash:1975:EfimamVfbt,Bordag:2009:ACE} the frequencies $\omega_j$ are complex numbers for dissipate media such as the conductive surfaces under consideration and formula (\ref{eq:EPSDef}) has no physical meaning. At the same time they noted that the formula of Lifshitz type (\ref{eq:EtetmPlanar}) is still valid. Last consideration in Ref. \cite{Nesterenko:2012:Lfbassm} shows that this formula contains contributions of surface (boundary) plasmons and scattering states. Hereinafter we will use the Lifshitz expression (\ref{eq:EtetmPlanar}) for planar and spherical symmetries.

To find the zero-point energies $\omega_{j}^{\TM,\TE}$, as well $\Psi_{\TM,\TE}$, we consider the electric and magnetic fields boundary conditions for the planar system. The electromagnetic vacuum fields are 
\begin{equation}
\bm{E} = \bm{e}(z) e^{ik_x x+ ik_y y - i \omega t}, \ \bm{H} = \bm{h}(z) e^{ik_x x+ ik_y y - i \omega t},
\end{equation}
where vectors $\bm{e}(z), \bm{h}(z)$ are the fields amplitudes, which depend on the $z$ coordinate only. 

The conductivity is responsible for an induced surface current density in each layer, $\bm{j}_s = \sigma \bm{E}$, which must be reflected in the TM and TE boundary conditions.
All TM field components are expressed   in terms of the single $e_z$ component, which satisfies the wave equation,
\begin{eqnarray}
e_x &=& \frac{i k_x}{k_\perp^2} e'_z,\   h_x =  +\frac{\omega k_y}{c k_\perp^2} e_z, \ann
e_y &=& \frac{i k_y}{k_\perp^2} e'_z,\   h_y =  -\frac{\omega k_x}{c k_\perp^2} e_z,
\end{eqnarray}
where $k_\perp^2 = k_x^2 + k_y^2$ and $k^2 = k_\perp^2 - \omega^2/c^2$ and prime denotes derivative with respect to $z$.  
All TE field components are expressed   in terms of the single component $h_z$, which also satisfies the wave equation,
\begin{eqnarray}
e_x &=& - \frac{\omega k_y}{c k_\perp^2} h_z,\   h_x = \frac{i k_x}{k_\perp^2} h'_z, \ann
e_y &=& + \frac{\omega k_x}{c k_\perp^2} h_z,\   h_y = \frac{i k_y}{k_\perp^2} h'_z.
\end{eqnarray}

Thus the appropriate boundary conditions for the TM modes are 
\bs\label{eq:PlanarBCTETM}
\begin{eqnarray}
[e'_z]_0 &=& 0, \ [e'_z]_d = 0, \ann{}
[e_z]_0 &=& -\frac{4\pi i \sigma}{\omega} e'_z|_0, \ann{}
[e_z]_d &=& -\frac{4\pi i \sigma}{\omega} e'_z|_d, \ann{}
e'_z|_{+L} &=& 0,\ e'_z|_{-L} = 0, \label{eq:PlanarBCTM}
\end{eqnarray}

The boundary conditions for the TE modes are
\begin{eqnarray}
[h_z]_0 &=& 0, \ [h_z]_d = 0, \ann{}
[h'_z]_0 &=& \frac{4\pi i \sigma \omega}{c^2} h_z|_0, \ann{}
[h'_z]_d &=& \frac{4\pi i \sigma \omega}{c^2} h_z|_d, \ann{}
h_z|_{+L} &=& 0,\ h_z|_{-L} = 0, \label{eq:PlanarBCTE}
\end{eqnarray}
\es
Here $[f(z)]_a = f(a-0)-f(a+0)$ at $z=a$. 

Solving the system of equations (\ref{eq:PlanarBCTETM}) enables obtaining the photon spectra for each mode type: 
\begin{gather}
\Psi_{\TM} (\omega)= \frac{k^2}{\epsilon_d^{2} \epsilon_L^{2}} \left\{ \frac{\omega ^2}{c^2} \epsilon _d^2 \left(\epsilon_L^4-1\right) - i \eta  k  \frac{\omega}{c}  \left(\epsilon_d^2 \left(\left(\epsilon _d^2+2\right) \epsilon _L^2\right.\right.\right.\ann
\left.\left.\left.+2 \epsilon_L^4+2\right)+\epsilon _L^2\right)  - \eta ^2 k^2 \left(\epsilon _d^2-1\right) \left(\epsilon _L^2+1\right)  \left(\epsilon _d^2+\epsilon _L^2\right)\right\}, \ann
\Psi_{\TE} (\omega)\! =\!\! \frac{1}{\epsilon _d^2\epsilon _L^2}\!\!\left\{\eta ^2 \frac{\omega^2}{c^2} \left(\epsilon _d^2-1\right) \left(\epsilon _L^2-1\right)  \left(\epsilon _d^2-\epsilon _L^2\right)-  i \eta  k   \frac{\omega}{c} \right.\ann
\left.   \times\left(\epsilon_d^2 \left(\left(\epsilon _d^2+2\right) \epsilon _L^2\!-\!2 \epsilon_L^4\!-\!2\right)\!+\!\epsilon _L^2\right)\!   + k^2 \epsilon _d^2 \left(\epsilon_L^4-1\right) \right\},\label{eq:Psi}
\end{gather}
where $\epsilon_L = e^{-Lk}, \epsilon_d = e^{-dk}, \eta = 2\pi \sigma/c$. 

Next, the limit $\lim_{d\to\infty}\lim_{L\to\infty} \mathcal{E}_{\TM,\TE}^{p}(s)$ is found using Eq. (\ref{eq:Psi}) and subtracted from $\mathcal{E}_{\TM,\TE}^{p}(s)$. Therefore, we obtain the renormalized  functions
\begin{eqnarray*}
\Psi_{\TM}^{\ren} (i\lambda c) &=& 1 - \rho_\TM^2 e^{-2 \kappa d},\ann
\Psi_{\TE}^{\ren} (i\lambda c) &=& 1 - \rho_\TE^2 e^{-2 \kappa d},
\end{eqnarray*}
where 
\begin{equation}
\rho_\TM = \frac{\eta \kappa}{\eta \kappa + \lambda},\ \rho_\TE = \frac{\eta \lambda}{\kappa + \eta \lambda}, 
\end{equation}
and $\kappa^2 = k_\perp^2 + \lambda^2$. Finally, after taking the $s\to 0$ limit, one finds 
\begin{eqnarray}
\mathcal{E}_{\TM}^{p} &=& \frac{\hbar c}{2\pi} \iint \frac{d^2 k_\perp}{(2\pi)^2} \int_0^\infty  \ln \left(1 - \rho_\TM^2 e^{-2 \kappa d} \right)d\lambda,\ann
\mathcal{E}_{\TE}^{p} &=&\frac{\hbar c}{2\pi} \iint \frac{d^2 k_\perp}{(2\pi)^2} \int_0^\infty  \ln \left(1 - \rho_\TE^2 e^{-2 \kappa d} \right)d\lambda.  
\end{eqnarray}

Using the spherical coordinates ($\theta \in (0,\pi/2), \varphi \in (0,2\pi)$) and setting $\kappa = y/d$ we obtain
\begin{equation}
\mathcal{E}_{\TM,\TE}^{p} = \frac{\hbar c }{d^3}Q_{\TM,\TE}^p(\eta),\label{eq:ETMTE}
\end{equation}
\begin{eqnarray*}
Q_{\TM}^p &=& \frac{1}{4\pi^2}\int_0^\infty  y^{2} dy\int_0^1 dx   \ln \left(1 - \frac{\eta^2}{(\eta +  x)^2}  e^{-2 y}\right),\ann
Q_{\TE}^p &=& \frac{1}{4\pi^2}\int_0^\infty  y^{2} dy\int_0^1 dx   \ln \left(1 - \frac{\eta^2 x^2}{(1 + \eta  x)^2}  e^{-2 y} \right),
\end{eqnarray*}
where $x=\cos\theta$.

Eq. (\ref{eq:ETMTE}) shows that the the Casimir energy has a distance dependence typical for metallic half-space systems. The characteristic constants $\hbar$ and $c$ are also extracted. The strength of the conductivity via $\eta=2\pi\sigma/c$ now determines the magnitude and sign of the Casimir interaction.

We consider the case of infinitely conducting planar sheets, first. It is easy to calculate that in this limit
\begin{equation}
\lim_{\eta \to \infty}Q_{\TM}^p = \lim_{\eta \to \infty}Q_{\TE}^p = -\frac{\pi^2}{1440}.\label{eq:Limits}
\end{equation}
which shows that the TE and TM modes contribute equally to the energy. Thus the well-known result for the Casimir interaction between infinitely conducting planes is recovered 
\begin{equation}
\lim_{\eta\to\infty} \mathcal{E}^{p} = - \frac{\pi^2}{720}\frac{\hbar c }{d^3}. 
\end{equation}

The $\eta\to 0$ limit is also examined. We find that
\begin{eqnarray}
\lim_{\eta\to 0}Q_{\TM}^p &=& -\frac{\eta}{4\pi^2}\left(-\frac{\pi^2}{6} - \frac{\pi^4}{360} + 4\ln 2 - \frac{1}{2} \zeta_R(3)\right),\ann 
\lim_{\eta \to 0}Q_{\TE}^p &=& -\frac{\eta^2}{48 \pi^2}, \label{eq:LimQpl}
\end{eqnarray} 
where $\zeta_R(s)$ is the Riemann zeta-function. Eq. (\ref{eq:LimQpl}) shows that both contributions are negative. Also, the linear dependence on $\eta$ shows that the TM contribution dominates the energy. 

In addition to the above limits, the graphene/graphene Casimir energy can also be calculated. Using the universal graphene conductivity $\sigma_{gr}= e^2/4\hbar, \eta_{gr}=\pi\alpha/2=0.0114\ll 1$ ($\alpha$ is fine-structure constant), one obtains from Eq. (\ref{eq:ETMTE}), 
\begin{equation}
\mathcal{E}^{p} = -  \frac{e^2}{32\pi d^3} Z,
 \end{equation} 
where $Z =4\left(-\frac{\pi^2}{6} - \frac{\pi^4}{360} + 4 \ln 2 - \frac{1}{2} \zeta_R (3)\right) \approx 1.024$. This result is consistent with a previously obtained Casimir energy via other methods  \cite{Drosdoff:2010:Cfgs} for which $Z=1$.   

\begin{figure}[tbh]
\centerline{\includegraphics[width=8cm]{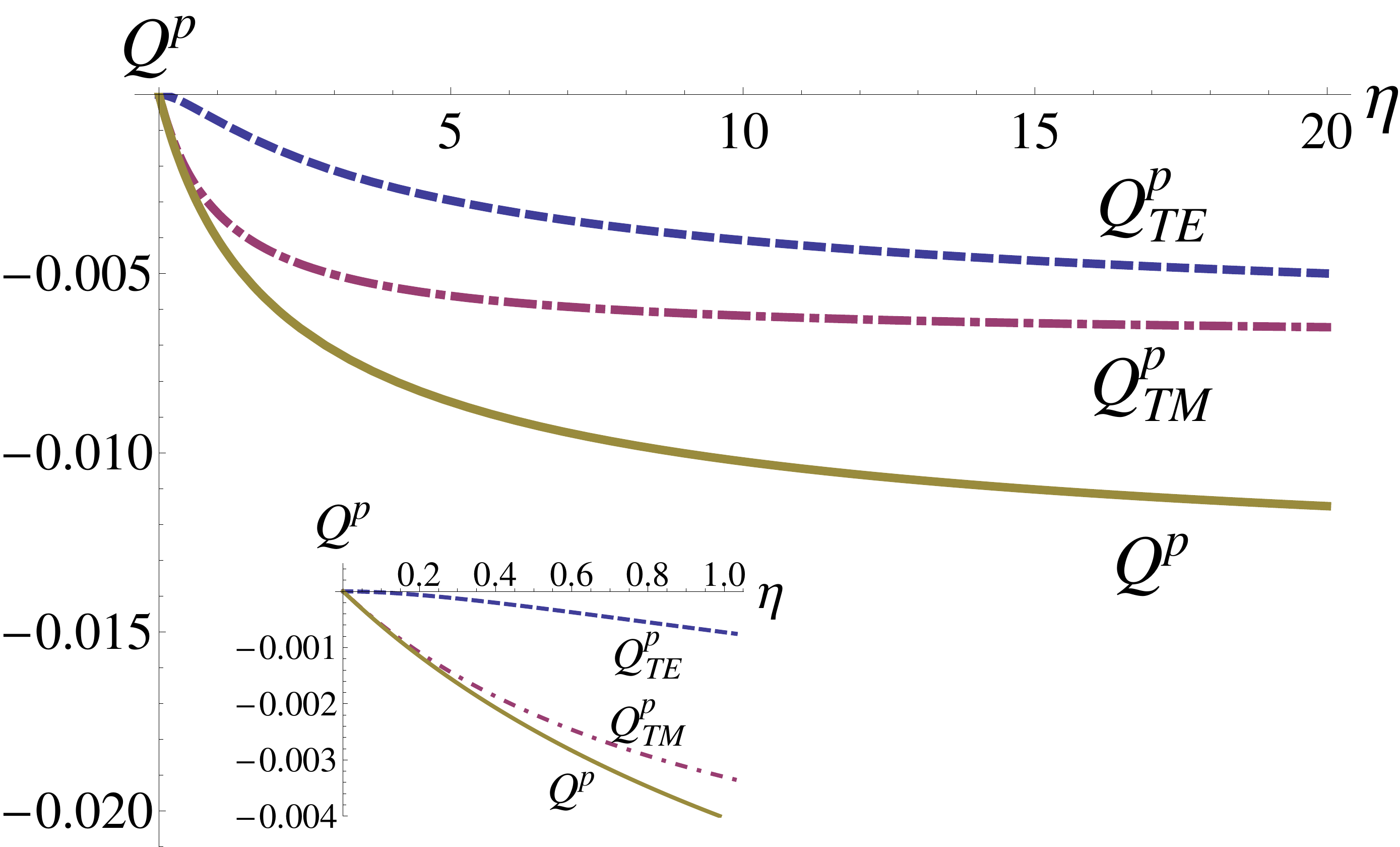}}
\caption{$Q_{\TE}^p$, $Q_{\TM}^p$ and their sum $Q_{\TM}^p + Q_{\TE}^p$ (solid line) as a function of $\eta$. The insert shows that for small $\eta \ll 1$ the function $Q_{\TM}^p \sim \eta$ whereas $Q_{\TE}^p \sim \eta^2$ and the main contribution comes from the TM mode. For the ideal conductor $\eta \to \infty$ both modes have the same contributions.} \label{fig:pssum}
\end{figure}

We also investigate how the intermediate region of $\sigma$ affects the interaction.  Fig. \ref{fig:pssum} shows the $Q_{\TE,\TM}^p$ behavior as a function of $\eta$. At small $\eta$ the majority of the contribution is attributed to the TM modes. As $\eta$ increases, the TE modes role becomes more prominent. Finally, at $\eta\to\infty$ both modes contribute equally to the interaction. 

The Casimir energy for thin materials has been considered previously in Ref. \cite{Parashar:2012:EsdpCiebpitp}, where the authors employ $\delta$-function boundary conditions. The derived Green's function dyadic for thin plates enables one to include dielectric and magnetic properties in terms of anisotropic response functions. In this work, we choose the more natural route by considering the 2D conductivity of the involved materials. 

\section{Spherical symmetry} \label{Sec:SpSym}

The Casimir energy for an infinitely thin spherical shell with radius $R$ and constant conductivity is considered next. The mode summation approach for perfectly conductive sphere was first considered in Ref.  \cite{Nesterenko:1998:SmfctCefas}. The general expression for $\mathcal{E}_{\TM,\TE}^s $ via the zero-point energy summation in Eq. (\ref{eq:EtetmPlanar})  is still valid. The electromagnetic field in the spherical geometry can be represented as TE and TM modes similar to the planar case. The TE modes, however, are characterized by the radial component of the magnetic field, while the TM modes are determined by the radial component of the electric field. The electric and magnetic fields are related to the radial function 
\begin{equation}
f(r) = a j_l (kr) + b y_l (kr),
\end{equation}
where $a,b$ are coefficients and $k=\omega /c$. Also, $ j_l (kr)$, $y_l (kr)$ are the spherical Bessel functions of first and second kind, respectively.

Similar to the planar case, the spherical shell is placed in a larger concentric shell with radius $L>R$ with Dirichlet boundary conditions. The $L\to\infty$ limit is taken in the final expression for the energy. The boundary conditions are found as follows:
\bs\label{eq:BCSum}
\begin{gather}
\mathsf{TE} : [rf]_R=0, [(rf)']_R = \frac{4\pi i \sigma \omega}{c^2} (rf), f|_L =0, \adb \label{eq:BC0}\\
\mathsf{TM} : [(rf)']_R=0, [rf]_R = -\frac{4\pi i \sigma}{\omega} (rf)', f|_L =0. \label{eq:BC}
\end{gather} 
\es
Similar expressions hold for a plasma shell with the difference that the constant conductivity $\sigma$ is replaced by the hydrodynamic conductivity $\sigma_h=- i ne^2/(m\omega)$, where $n$ is the electron density \cite{Barton:2004:Cesps,*Barton:2005:CefpsIE,Bordag:2008:Ovesps,*Khusnutdinov:2011:vdWibaaaasps,*Khusnutdinov:2012:tCiasps}.

\subsection{The energy due to TE modes}
Using Eqs. (\ref{eq:BC0}), the TE mode spectrum is
\begin{eqnarray}
\Psi_{\TE}(ik) &=& y^{-l-1} \left\{ s_l (y) - 2 \eta s_l (x)\right.\ann
&\times& \left.\left[s_l (x) e_l (y) -  s_l (y) e_l (x) \right]\right\}, 
\end{eqnarray}
where $y = kL, x = kR$.  Also, 
\begin{equation}
s_l(x) = \sqrt{\frac{\pi x}{2}} I_{l+1/2}(x),\ e_l(x) = \sqrt{\frac{2 x}{\pi}} K_{l+1/2}(x), 
\end{equation}
and $I_\nu, K_\nu$ are the modified Bessel functions of first and second kind, respectively. 

The $L\to\infty$ limit is now taken. Due to the asymptotic behavior of the Bessel functions in this case and omitting the Minkowski contribution, it is found that $\Psi_{\TE}(ik)|_{L\to\infty}$ makes no contribution. The remaining part is written as
\begin{equation}
f_{\TE}(ik) =  1 + 2 \eta s_l (x)e_l (x) = 1 + 2 \eta x I_\nu (x)K_\nu (x). \label{eq:fte}
\end{equation}
Further, in the framework of the zeta-regularization approach for spherical geometry, using Eq. (\ref{eq:fte}) the TE Casimir energy is found
\begin{eqnarray}
\mathcal{E}_{\TE}^s (s) &=& -\hbar c \Lambda^{2s} \frac{\cos \pi s}{\pi } R^{2s -1} \sum_{l=1}^\infty \nu^{2-2s} \ann
&\times&\int_0^\infty dz z^{1-2s} \frac{\partial}{\partial z}    \ln f_{\TE} (\nu z), \label{eq:ETEs}
\end{eqnarray}
where $\nu=l+1/2$. The calculations are further advanced by writing the energy in the form
\begin{equation}
\mathcal{E}_{\TE}^s(s) = \left\{\mathcal{E}_{\TE}^s(s) - \mathcal{E}^{s,\as}_{\TE}(s)\right\} + \mathcal{E}^{s,\as}_{\TE}(s).
\end{equation}
The added and subtracted term is calculated via the Debye uniform expansion of the modified Bessel functions \cite{Abramowitz:1970:eHMFFGMT}, for which the integrand becomes
\bs\label{eq:fteas}
\begin{gather}
\frac{\partial}{\partial z}\ln f_{\TE}^{\as} (\nu z) = \sum_{n\geq 0} \frac{F^{\TE}_{2n}(z,\eta)}{\nu^{2n}},
\end{gather}
where
\begin{eqnarray}
F^{\TE}_0 (z,\eta) &=& \frac{\eta t^3}{1+\eta t z},\adb\\
F^{\TE}_2 (z,\eta) &=& - \frac{\eta t^3}{8 (1+ \eta t z)^2} \left\{ 2 - 27 t^2 +  60 t^4 - 35 t^6\right.\ann
&+& \left. 2 t^3 z^3\eta (1 - 12 t^2 + 15 t^4)\right\},
\end{eqnarray}
\es
and $t = 1/\sqrt{1+z^2}$. Note that for the renormalization only the $F^{\TE}_0$ and $F^{\TE}_2$ are needed since the rest of the terms will give finite contributions. 

Taking the limit of $s\to 0$, we find that
\bs\label{eq:PSETEnum}
\begin{equation}
\mathcal{E}_{\TE}^{s,\num} = \lim_{s\to 0} \left\{\mathcal{E}_{\TE}^s(s) - \mathcal{E}^{s,\as}_{s,\TE}(s)\right\} = \frac{\hbar c}{R}Q_{\TE}^{s,\num}(\eta),
\end{equation}
where
\begin{gather}
Q_{\TE}^{s,\num}(\eta) =  - \sum_{l=1}^\infty \nu^{2} \int_0^\infty dz  z \left\{ \frac{ 2 \eta \left ( s_l(\nu z) e_l (\nu z) \right)'_z}{1+2\eta s_l(\nu z) e_l (\nu z)}\right.\ann
\left. - F^{\TE}_0(z,\eta) - \frac{1}{\nu^2} F^{\TE}_2(z,\eta) \right\}.
\end{gather}
\es
Thus the Casimir energy is expressed to contain the characteristic constants $\hbar$ and $c$ and the spherical radius explicitly. $Q_{\TE}^{s,\num}(\eta)$ is a function of the renormalized conductivity $\eta$ only and it is usually computed numerically.

The contribution from the Debye expansion can be evaluated analytically. Relevant details are provided in Appendix A. We find that 
\begin{eqnarray}
\mathcal{E}_{\TE}^{s,\as}(s) &=& - \hbar c\frac{\cos\pi s}{\pi R}\Lambda^{2s} \int_0^\infty dz z^{1-2s} \left\{ \zeta_H\left(2s-2,\frac 32\right)\right.\ann 
&\times& \left.F^{\TE}_0(z,\eta) - \zeta_H\left(2s,\frac 32\right)F^{\TE}_2(z,\eta) \right\},\label{eq:PSETEas-1}
\end{eqnarray}
where $\zeta_H (s,a)$ is the Hurwitz zeta function \cite{Bateman:1953:Htf-1} which has a singe simple pole at $s=1$. Appendix A shows that the analytic contribution in the energy has no poles in terms of $s$. It is finite at $s=0$ and it is found in the following form:
\bs\label{eq:PSETEas}
\begin{equation}
\mathcal{E}_{\TE}^{s,\as}(0) = \frac{\hbar c}{R}Q_{\TE}^{s,\as}(\eta),
\end{equation}
where 
\begin{eqnarray}
Q_{\TE}^{s,\as}(\eta) &=& \frac{17}{128} - \frac{1}{12 \pi \eta} + \frac{3}{32\eta^2} + \frac{5}{8\pi \eta^3} - \frac{5}{16\eta^4}  \ann 
&+& \frac{ (5 - 4 \eta^2 - 2\eta^4)}{8\pi \eta^4 \sqrt{\eta^2 -1}}\ln(\eta + \sqrt{\eta^2 -1}).
\end{eqnarray}
\es

Thus Casimir energy from the TE modes 
\begin{equation}
\mathcal{E}_{\TE}^s = \frac{\hbar c}{R}Q_{\TE}^s(\eta),
\end{equation}
where $Q_{\TE}^s = Q_{\TE}^{s,\as} + Q_{\TE}^{s,\num}$ is a sum of the numerical and the analytic contributions.  Using the derived expressions in Eqs. (\ref{eq:PSETEnum}),(\ref{eq:PSETEas}), the  $\eta\to 0$ and $\eta\to\infty$ limiting cases can be studied. We find that
\begin{gather}
\eta \to 0 \ann 
Q_{\TE}^{s,\as} = \frac{\eta}{4\pi}, Q_{\TE}^{s,\num} =  O(\eta^2) , Q_{\TE}^s =\frac{\eta}{4\pi}, \ann
\eta \to \infty\ann
Q_{\TE}^{s,\as} = \frac{17}{128}, Q_{\TE}^{s,\num} = 0.0009 , Q_{\TE}^s =  0.1337.\label{eq:LimTE}
\end{gather}
These results show that $Q_{\TE}^{s,\as}\gg Q_{\TE}^{s,\num}$ in all cases. The dominating contribution originates from the analytic (Debye expansion) part, which is $\sim \eta$ for small conductivity and it is constant, for an infinitely conducting shell. Another interesting feature is that $Q_{\TE}^s>0$. This indicates that unlike in the two parallel planes, the Casimir energy due to these modes is repulsive.  

\begin{figure}[ht]
\centerline{\includegraphics[width=8cm]{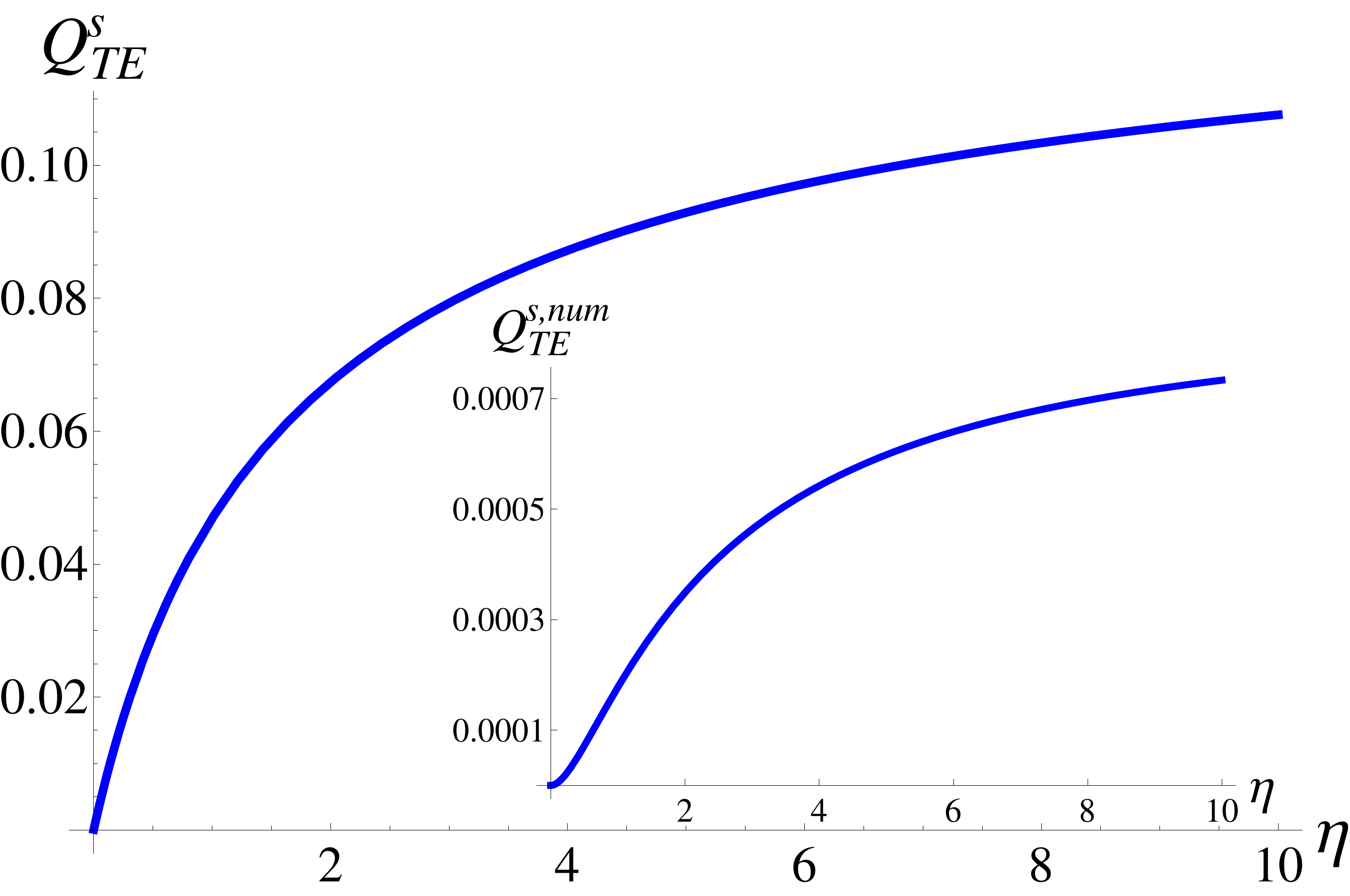}}
\caption{$Q_{\TE}^s$ as a function of $\eta$. The insert shows the much smaller contribution from the numerical $Q_{\TE}^{s,\num} \ll Q_{\TE}^s$. $Q_{\TE}^s$ is always positive.}\label{fig:QTE}
\end{figure}

Fig. \ref{fig:QTE} shows the computation  $Q_{\TE}^s = Q_{\TE}^{s,\as} + Q_{\TE}^{s,\num}$ for intermediate values of $\eta$. It is clear that the Casimir energy is completely determined by $Q_{\TE}^{s,\as}$ for all values of the conductivity.

\subsection{The energy due to TM modes} 

The TM mode energy spectrum is found using the boundary conditions from Eq. (\ref{eq:BC}) 
\begin{equation*}
\Psi_{\TM}(ik) = y^{-l} \{ s_l (y) + 2 \eta  s'_l (x)\left( s'_l (x) e_l (y) -  s_l (y) e'_l (x) \right)\}, 
\end{equation*}
where $y = kL, x = kR$ and $k=\omega/c$. After taking the limit of $L\to\infty$ and omitting the Minkowski contribution $ y^{-l-1} s_l (y)$, it is obtained
\begin{gather}
f_{\TM}(ik) = x\left\{ 1 - 2 \eta  s'_l (x)e'_\nu (x)\right\}, 
\end{gather}
which is related to the Casimir energy in a similar manner as for the TE modes (Eq. (\ref{eq:ETEs}))
\begin{eqnarray*}
\mathcal{E}_{\TM}^s (s) &=& -\hbar c \Lambda^{2s} \frac{\cos \pi s}{\pi } R^{2s -1} \sum_{l=1}^\infty \nu^{2-2s} \ann
&\times&\int_0^\infty dz z^{1-2s} \frac{\partial}{\partial z}\ln f_{\TM} (\nu z).
\end{eqnarray*}

Adding and subtracting the Debye asymptotic expansion again, the TM Casimir energy is represented as
\begin{equation}
\mathcal{E}_{\TM}^s(s) = \left\{\mathcal{E}_{\TM}^s(s) - \mathcal{E}^{s,\as}_{\TM}(s)\right\} + \mathcal{E}^{s,\as}_{\TM}(s).
\end{equation}
The asymptotic Debye  expansion of $f_{\TM} (\nu z)$ has the following form 
\begin{equation}
f_{\TM}^{\as} (\nu z) = z  + \frac{\eta}{t} + t \eta \sum_{j\geq 1} \frac{c_j^{\TM}}{\nu^{2j}},
\end{equation}
where $c_j^{\TM}$ are polynomials in terms of $t = 1/\sqrt{1+z^2}$ of even degree and $ c_1^{\TM} = - \left(1-6t^2 + 7t^4\right)/8$. For the renormalization only  $c_1^{\TM}$ is needed as the rest of the terms give a finite contribution. 

For the part to be computed numerically after the $s\to 0$ limit, we find that 
\bs\label{eq:ETMnumFin}
\begin{equation}
\mathcal{E}_{\TM}^{s,\num} =  \lim_{s\to 0}  \left\{\mathcal{E}_{\TM}^s(s) - \mathcal{E}^{s,\as}_{\TM}(s)\right\} =   \frac{\hbar c}{R}Q_{\TM}^{s,\num}(\eta),
\end{equation}
where
\begin{gather}
Q_{\TM}^{s,\num} = -\frac{1}{\pi} \sum_{l=1}^\infty \nu^{2} \int_0^\infty dz \left( - 2 \left(1 + \frac{1 - \frac{1}{4\nu^2}}{z^2}\right)\right. \ann
\times \left.\frac{ \eta z (s_l(\nu z)e_l(\nu z))'_z }{1- 2\eta s'_l(\nu z) e'_l (\nu z)}+ \frac{\eta t}{z  + \frac{\eta}{t}} - \frac{\eta tz}{8 \nu^2 (z  + \frac{\eta}{t})^2}\right. \ann  
\times \left.  \{2\! -\! 25 t^2\! +\! 60 t^4\! -\!35 t^6\!\! +\!2\eta t z (1\!-\!12t^2\!\! +\! 21t^4)\} \right).
\end{gather}
\es

The contribution due to the Debye expansion can be evaluated analytically
\begin{gather}
\mathcal{E}_{\TM}^{s,\as}(s) = -\hbar c\frac{\cos\pi s}{\pi R}\Lambda^{2s} \sum_{l=1}^\infty \nu^{2-2s} \int_0^\infty dz z^{1-2s} \left\{ \frac{1+ t z\eta }{z  + \frac{\eta}{t}}\right.\ann
+ \left.\frac{\eta t}{8 \nu^2 (z  + \frac{\eta}{t})^2} \{2 - 25 t^2 + 60 t^4 -35 t^6\right.\ann
\left. +2\eta t z (1-12t^2 + 21t^4)\} \right\}.
\end{gather}
As shown in Appendix B, this contribution to the Casimir energy has no poles and it is finite at $s=0$. It has the following form
\bs\label{eq:ETMasFin}
\begin{equation}
\mathcal{E}_{\TM}^{s,\as}(0) = \frac{\hbar c}{R}Q_{\TM}^{s,\as}(\eta),
\end{equation}
where
\begin{eqnarray}
Q_{\TM}^{s,\as}(\eta) &=& -\frac{\eta}{96\pi } (4 (10-21\eta^2) - 3\pi \eta (9-14\eta^2) )\ann
&-& \frac{\eta^3}{8\pi} \frac{8-7\eta^2}{\sqrt{1-\eta^2}} \ln \frac{1+\sqrt{1-\eta^2}}{\eta}. 
\end{eqnarray}
\es

The expressions for $Q_{\TM}^{s,\num}$ and $Q_{\TM}^{s,\as}$ from Eqs. (\ref{eq:ETMnumFin}), (\ref{eq:ETMasFin}) can now be used to obtain the $\eta\to\infty$ and $\eta\to 0$ limiting cases for the Casimir energy from the TM mode contribution. These limiting cases are summarized as follows:
\begin{gather}
\eta \to 0\ann
Q_{\TM}^{s,\as} = - \frac{5\eta}{12\pi}, Q_{\TM}^{s,\num} = -0.00123 \eta , Q_{\TM}^s = -0.1338 \eta, \ann
\eta \to \infty\ann
Q_{\TM}^{s,\as} = - \frac{11}{128}, Q_{\TM}^{s,\num} = -0.0016 , Q_{\TM}^s = -0.0875.\label{eq:LimTM}
\end{gather}
Again, the numerical contribution plays a minor role to the interaction. The TM contribution is obtained to be negative. The dominant $Q_{\TM}^{s,\as}$ for an infinitely conducting spherical shell is in agreement with previous findings \cite{Bordag:2008:Ovesps}.

\begin{figure}[ht]
\centerline{\includegraphics[width=8cm]{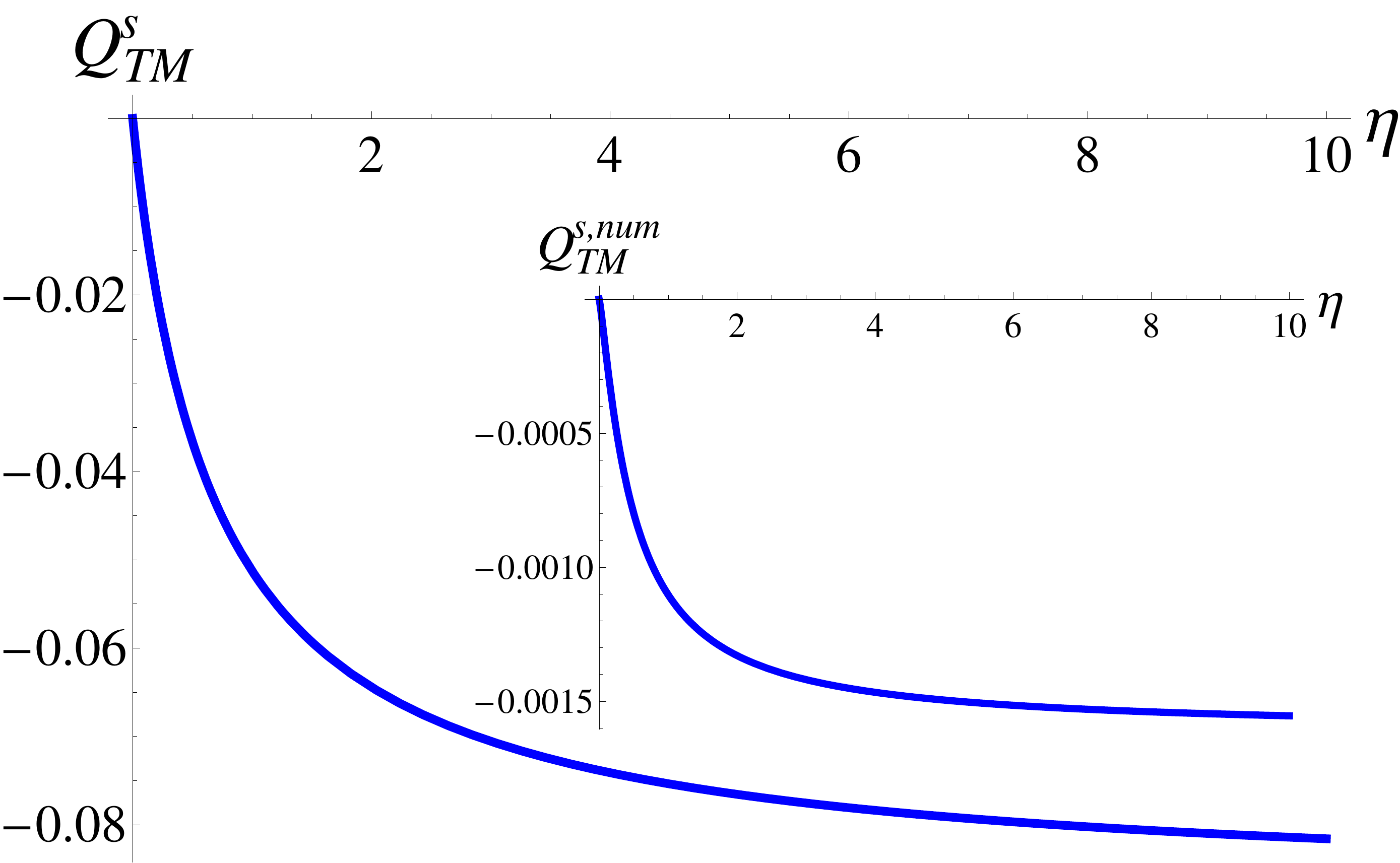}}
\caption{$Q_{\TM}^s$ as a function of $\eta$. The insert shows that the numerical part is much smaller as compared to the Debye expansion,  $Q_{\TM}^{s,\num} \ll Q_{\TM}^s$. The $Q_{\TM}^s$ function is always negative.}\label{fig:QTM}
\end{figure}

In Fig. \ref{fig:QTM}, results are shown for the intermediate range of $\eta$. It is noted that the $Q_{\TM}^s$ is negative and it is determined mainly by $Q_{\TM}^{s,\as}$ in the entire $\eta$ range.

\subsection{The total energy} 

The total Casimir energy due to both polarizations is examined next 
\begin{equation}
\mathcal{E}^s = \frac{\hbar c}{R}Q^s(\eta),
\end{equation}
where $Q^s = Q_{\TE}^s + Q_{\TM}^s$. For an ideal conductor $\eta\to\infty$, Eqs. (\ref{eq:LimTE}), (\ref{eq:LimTM}) yield
\begin{equation}
\lim_{\eta\to \infty} Q^s(\eta) = 0.046,
\end{equation}
in agreement with the classical result obtained by Boyer \cite{Boyer:1968:QEZEoaCSSatCMfaCP} (see also  different calculations in Ref. \cite{Bordag:2008:Ovesps}). Thus the energy is determined mainly by the TE contribution, which is responsible for the repulsive nature of the Casimir interaction of the spherical shell.

For a conductor with small conductivity $\eta\to 0$, Eqs. (\ref{eq:LimTE}), (\ref{eq:LimTM}) give 
\begin{equation}
Q^s(\eta) \approx   -0.0542 \eta,
\end{equation}
which indicates that the Casimir interaction is mainly determined by the TM modes resulting in an attractive force. 

\begin{figure}[ht]
\centerline{\includegraphics[width=8cm]{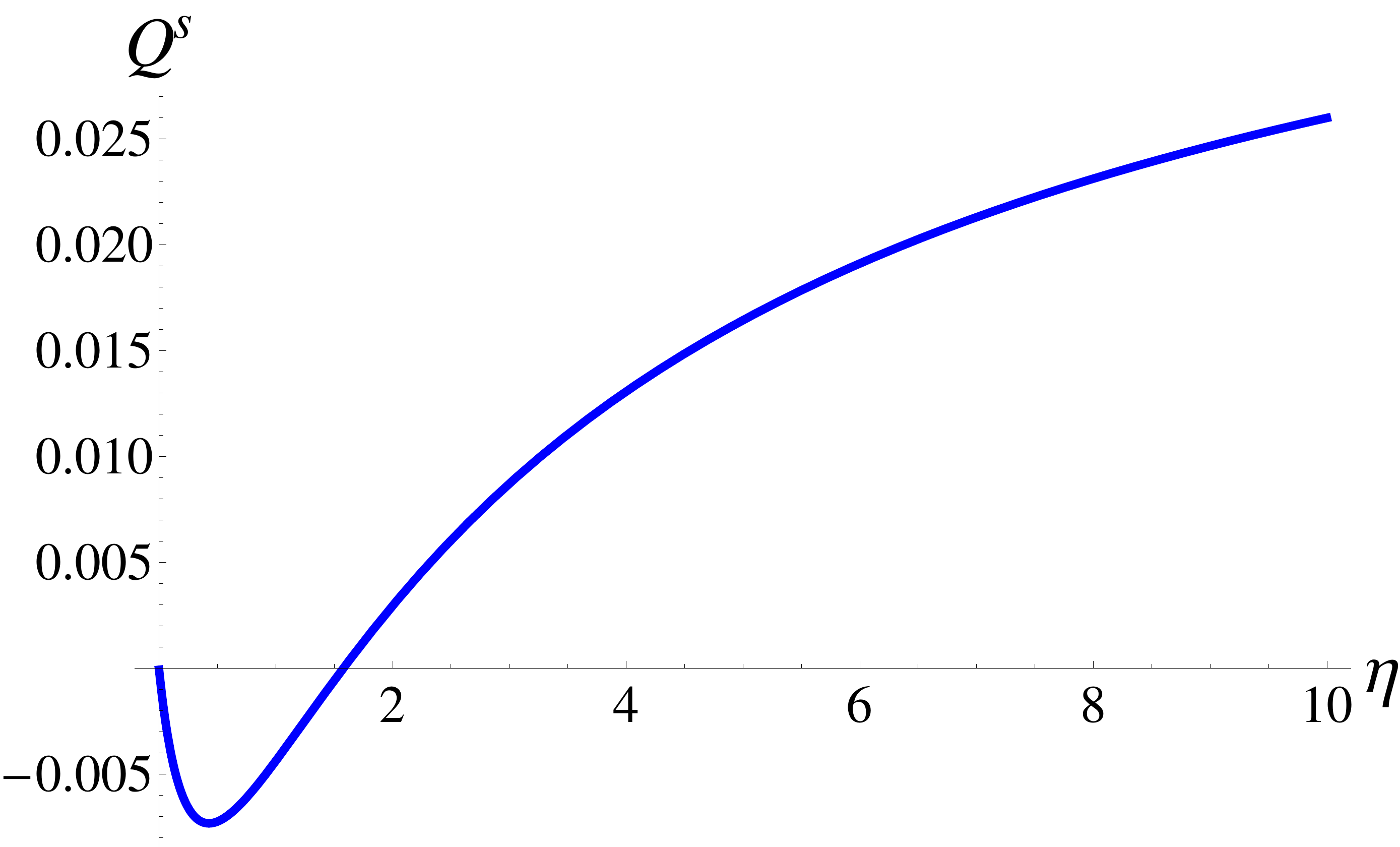}}
\caption{The total contribution  $Q^s = Q_{\TE}^s + Q_{\TM}^s$ vs $\eta$.  At the point $\eta_{cr} = 1.578 (\sigma_{cr} = 0.251c)$ the energy and the force are zero for arbitrary radii of the sphere. For the sphere with ideal conductivity $(\eta\to \infty)$ we obtain Boyer result $\lim_{\eta\to\infty}Q_s = 0.046$.}\label{fig:Q}
\end{figure}

Fig. \ref{fig:Q} shows how $Q^s$ evolves as a function of $\eta$. It is found that there is a critical point at $\eta_{cr}$, at which $Q^s$ becomes zero and it changes sign. We estimate that $\eta_{cr} = 1.578 (\sigma_{cr} = 0.251c)$. Thus for $\sigma < \sigma_{cr}$, the energy is negative and the Casimir force is attractive, while for $\sigma > \sigma_{cr}$ the energy is positive and the force is repulsive.

It is also interesting to find the Casimir energy assuming the spherical shell has the same conductivity as the universal conductivity of a planar graphene sheet -- $\sigma_{gr} = e^2/4\hbar$. Comparing  $\eta_{gr} = 2\pi \sigma_{gr}/c = 0.01146$ with the critical $\eta_{cr}$, it is clear that   $\eta_{gr} \ll \eta_{cr}$. Thus we may use the asymptotic expansion for $\eta \ll 1$ for the TE and TM contributions. From Eqs. (\ref{eq:LimTE}), (\ref{eq:LimTM}), we arrive at
\begin{equation*}
\mathcal{E}^s = - 0.000621\frac{\hbar c}{R} = -0.0542\frac{\hbar c }{R}\eta_{gr}  = -0.0851 \frac{e^{2}}{R}.
\end{equation*}
Therefore, the Casimir force for a spherical shell with conductivity equal to the one for graphene is attractive. Furthermore, the energy does not depend on the characteristic $\hbar$ and $c$ constants, which is similar to the case of two Casimir interaction between two planar graphenes. 

\section{Discussion and conclusions}\label{Sec:Disc}

The Casimir energy for surfaces characterized with a constant conductivity is studied via the generalized zeta function regularization procedure. In particular, the energy between two planar surfaces and the energy of a spherical shell are considered. The motivation for this investigation comes from many experimental works, which have demonstrated that graphene is characterized by a constant universal conductivity a relatively broad frequency range. Other experiments have shown that graphene is quite flexible and it can be folded into various shapes. The existence of such materials is an important stimulator to expand the use of regularization techniques for Casimir interactions. In this way, we further develop this approach and facilitate finding new feature in this ubiquitous force beyond specific materials.  

We are able to express the energy by containing characteristic constants, $\hbar$ and $c$, and distance dependence. The constant conductivity $\sigma$ is captured in a function of a dimensionless parameter $\eta=2\pi\sigma/c$. In particular, for two parallel surfaces the Casimir energy is expressed as $\mathcal{E}^p = \hbar c Q^{p}(\eta)/d^3$, while for a spherical shell the energy is  $\mathcal{E}^s= \hbar c Q^{s}(\eta)/R$. Therefore, the distance dependence of the interaction is the same as the one for perfect metallic systems in both cases.

Most of our efforts were directed towards calculating the $Q(\eta)$ function, which involved using the zeta-regularization procedure. It is known that for typical metals in the framework of this method divergences always occur, which are usually removed by appropriate  renormalization procedure for the parameters of the classical part of the energy (see Ref.  \cite{Bordag:1997:Cefmfitb},  and Ref. \cite{Bordag:2008:Ovesps} for plasma spherical shell). This procedure is based on using the heat kernel and heat kernel coefficients theory, which is well developed by now \cite{Vassilevich:2003:HkeUm}. The heat kernel coefficients strictly depend on the geometry of the  background and  boundary conditions. They have been calculated for simple boundary conditions. In the case of unusual boundary conditions the coefficients acquire complicated structure.  For example, for the spherical plasma shell \cite{Bordag:2008:Ovesps}, the heat kernel for this nontrivial boundary has additional logarithmic contribution. For the systems studied here, the spectral problem is formulated with non-standard boundary conditions for the planar and spherical symmetries.  Our calculations show that for constant $\sigma$ there is no divergent behavior in the Casimir energy. 

Further important implications for the Casimir interactions are found. For the planar case, the interaction is always attractive. The obtained analytic expressions show that in the small conductivity limit, the TM modes determine the interaction, while in the perfect metal limit, both TE and TM modes contribute equally. For the spherical case, the situation is quite different. Depending on the magnitude of $\sigma$, the Casimir interaction can be attractive or repulsive. In the small conductivity range, the TM modes overwhelm the TE contribution, and the Casimir interaction is attractive. In the large conductivity range, the situation is reversed. This implies that at a critical conductivity, the Casimir energy of the spherical shell is zero.  It is also interesting that in both geometries $Q\sim\eta$ in the $\eta\to 0$ limit. In addition, we are able to extract and evaluate analytically the dominant contributions to the energy. Previously obtained results for perfect conductors and parallel graphenes are also recovered.

The change in sign of the Casimir force was also shown to exist for a spherical shell with conductivity characterized by a frequency dependent plasma model \cite{Bordag:2008:Ovesps}. However, the attractive/repulsive transition is radius dependent: the interaction is attractive for smaller radii and repulsive for larger radii. We point out that for the case of $\sigma=const$, this attractive/repulsive transition is determined by the magnitude of the $\sigma$ and it is independent of the radius of the shell.

In conclusion, this investigation, expired by recent discoveries in surface materials, extends the applicability of an important tool for Casimir interaction calculations in terms of zeta-function regularization developments. Furthermore, the role of geometry together with constant conductivity response properties shows intriguing functionalities in this universal force. Further developments in terms of taking into account the role of temperature and the frequency dependence in the response properties of such systems will be investigated in the future. 

\appendix 
\section{Calculation of $\mathcal{E}_{\TE}^{s,\as}(s)$}\label{Sec:AppA}

The TE energy $\mathcal{E}_{\TE}^{\as}(s)$ is calculated using Eq. (\ref{eq:PSETEas-1}) by defining the following integrals 
\begin{eqnarray}
N_k(s)&=& \eta^2 \cos \pi s  \int_0^\infty dz \frac{z^{4-2s}t^{6+2k}}{(1+\eta t z)^2}, -\frac{1}{2}-k < \Re s <\frac{5}{2},\ann
&&\ann
M_k(s)&=& \eta \cos \pi s  \int_0^\infty dz \frac{z^{1-2s}t^{3+2k}}{(1+\eta t z)^2}, -\frac{1}{2} -k < \Re s <1,\ann
J_0(s) &=& \eta \cos\pi s\int_0^\infty dz \frac{z^{1-2s}t^3}{1+\eta t z}, -\frac{1}{2} < \Re s <1 . \label{eq:Ints}
\end{eqnarray}
where $k\ge 0, t=1/\sqrt{1+z^2}$. Within the specified in  Eq. (\ref{eq:Ints}) domains of convergence, the energy becomes
\begin{gather*}
\mathcal{E}_{\TE}^{s,\as}(s) = -\frac{\Lambda^{2s}}{\pi R} \left\{J_0(s)\zeta_H\left(2s-2,\frac 32\right) -  \frac{\zeta_H(2s,\frac 32)}{8}\right. \ann
\left. \times \left[ 2M_0(s,\eta) - 27 M_1(s,\eta) +  60 M_2(s,\eta) - 35 M_3(s,\eta)\right]\right. \ann
+ \left. \frac{\zeta_H(2s,\frac 32)}{4} \left[N_0(s,\eta) - 12 N_1(s,\eta) + 15 N_2(s,\eta)\right]\right\}.
\end{gather*}
Making the analytic continuation in the point $s=0$, it is found that all expressions are finite as shown
\begin{widetext}
\begin{eqnarray}
J_0(0,\eta) &=&  \frac{\pi}{2} - \frac{\ln (\eta + \sqrt{\eta^2 -1})}{\sqrt{\eta^2 -1}}, \ann
M_0(0,\eta) &=& \frac{\eta}{1-\eta^2} + \frac{\eta^2}{(\eta^2-1)^{3/2}}\ln(\eta + \sqrt{\eta^2 -1}),\ann
M_1(0,\eta) &=& -\frac{2 }{\eta} + \frac{\pi }{\eta^2 } + \frac{ \eta^2-2}{\eta^2\sqrt{\eta^2-1}}\ln(\eta + \sqrt{\eta^2 -1}),\ann
M_2(0,\eta) &=& - \frac{7 }{3 \eta} + \frac{3\pi}{2\eta^2} + \frac{4}{\eta^3} - \frac{2\pi}{\eta^4} + \frac{\sqrt{\eta^2 - 1}(\eta^2 - 4)}{\eta^4 }\ln(\eta + \sqrt{\eta^2 -1}),\ann
M_3(0,\eta) &=& -\frac{38}{15 \eta} + \frac{15\pi}{8\eta^2} + \frac{9}{\eta^3} - \frac{5\pi}{\eta^4} - \frac{6}{\eta^5} + \frac{3\pi}{\eta^6} +\frac{(\eta^2 - 1)^{3/2}(\eta^2 - 6)}{\eta^6 }\ln(\eta + \sqrt{\eta^2 -1}) \ann
N_0(0,\eta) &=& -\frac{3-2\eta^2 }{\eta (1-\eta^2)} + \frac{\pi (6+\eta^2) }{4 \eta^2} + \frac{ (3 -4 \eta^2)\sqrt{\eta^2 -1}}{\eta^2 (1-\eta^2)^{2}}\ln(\eta + \sqrt{\eta^2 -1}),\ann
N_1(0,\eta) &=&  \frac{\pi}{16} -\frac{2 }{3\eta} +\frac{3\pi}{4\eta^2} + \frac{5}{\eta^3} - \frac{5\pi}{2\eta^4}  - \frac{ (5 - 4 \eta^2)\sqrt{\eta^2 -1}}{\eta^4 (1-\eta^2)}\ln(\eta + \sqrt{\eta^2 -1}),\ann
N_2(0,\eta) &=&  \frac{\pi}{32} -\frac{2 }{5\eta} +\frac{9\pi}{16\eta^2} + \frac{19}{3\eta^3} - \frac{15\pi}{4\eta^4} - \frac{7}{\eta^5} + \frac{7\pi}{2\eta^6} + \frac{ (7 - 4 \eta^2)\sqrt{\eta^2 -1}}{\eta^6 }\ln(\eta + \sqrt{\eta^2 -1}).
\end{eqnarray}
\end{widetext}

The Hurwitz zeta function $\zeta_H(s,a)$ \cite{Bateman:1953:Htf-1} has a simple single pole at $s=1$,
\begin{equation}
\zeta_H(s,a) \approx \frac{1}{s-1}.
\end{equation}
In the points of interest for $\mathcal{E}_{\TE}^{\as}(s)$, the function is finite: $\zeta_H\left(-2,\frac 32\right) = -\frac{1}{4},\zeta_H\left(0,\frac 32\right) =-1$. 

As a result of the above expression, we obtain that the asymptotic TE energy can be written in the following form:
\begin{equation}
\mathcal{E}_{\TE}^{s,\as}(0) = \frac{Q_{\TE}^{s,\as}(\eta)}{R},
\end{equation}
where
\begin{eqnarray}
Q_{\TE}^{s,\as}(\eta) &=& \frac{17}{128} - \frac{1}{12 \pi \eta} + \frac{3}{32\eta^2} + \frac{5}{8\pi \eta^3} - \frac{5}{16\eta^4}  \ann 
&+& \frac{ (5 - 4 \eta^2 - 2\eta^4)}{8\pi \eta^4 \sqrt{\eta^2 -1}}\ln(\eta + \sqrt{\eta^2 -1}).
\end{eqnarray}

\section{Calculation of $\mathcal{E}_{\TM}^{s,\as}(s)$}\label{Sec:AppB}

The TM energy $\mathcal{E}_{\TM}^{\as}(s)$ is calculated by defining the following integrals with $(k\geq 0, t= 1/\sqrt{1+z^2})$
\begin{eqnarray}
B_k(s) &=& \eta^2 \cos\pi s \int_0^\infty \frac{z^{2-2 s} t^{2+2k}}{(z+\frac \eta t)^2} dz, -\frac{1}{2} - k < \Re z <\frac{3}{2},\ann
A_k(s) &=& \eta \cos\pi s \int_0^\infty \frac{z^{1-2 s} t^{1+2k}}{(z+\frac \eta t)^2} dz, -\frac{3}{2} - k < \Re z <1,\ann
I_0(s) &=& \cos\pi s \int_0^\infty \frac{z^{1-2 s} (1+ t z\eta)}{z+\frac \eta t} dz, \frac{1}{2} < \Re z <1.
\end{eqnarray}
Within the specified domains of convergence in the above equation, the energy becomes
\begin{gather}
\mathcal{E}_{\TM}^{s,\as}(s) = -\frac{\Lambda^{2s}}{\pi R} \left\{I_0(s) \zeta_H\left(2s-2,\frac 32\right) + \zeta_H\left(2s,\frac 32\right)\right.\ann
\times\left. \frac{1}{8} \left[2 A_0(s) - 25 A_1(s) + 60 A_2(s) -35 A_3(s)\right] \right. \ann
+ \left. \frac 14 \left[B_0(s)-12 B_1(s) + 21B_2(s)\right]\zeta_H\left(2s,\frac 32\right) \right\}. 
\end{gather}
After making the analytic continuation for $s=0$ and realizing that the Hurwitz zeta function is $\zeta_H\left(-2,\frac 32\right) = -\frac{1}{4},\zeta_H\left(0,\frac 32\right) =-1$, it is found that the $A_k$ and $B_k$ integrals are finite at $s=0$. 

Let us consider the integral $I_0$. It is convergent for $1/2 < \Re s <1$ and because of the denominator $z+\eta \sqrt{1+z^{2}}$, it has no zeros in the complex plane of $z$. One notes that due to the $\sqrt{1+z^2}$, the integrand has two branch points at $z=\pm i$. Changing the variable $z\to y = 1/z$ and extending the integration over the entire Re axes, we obtain
\begin{eqnarray}
I_0(s) = \frac{e^{-i\pi s}}{2} \int_{-\infty}^{+\infty} dy y^{2s-2} \frac{1 + \frac{\eta}{\sqrt{y^2+1}}}{1+\eta \sqrt{y^2+1}}.
\end{eqnarray}
After shifting the contour to the Im axes we arrive at the expression
\begin{equation}
I_0(s) = - \eta \int_1^\infty \frac{x^{2s}}{\sqrt{x^2 -1}} \frac{dx}{1+ \eta^2 (x^2 -1)}. 
\end{equation}
This integral is convergent for all $\Re s<1$. Thus at $s=0$, we find 
\begin{widetext}
\begin{eqnarray}
I_0(0) &=& -\frac{\eta}{\sqrt{1-\eta^2}} \ln \frac{1+\sqrt{1-\eta^2}}{\eta}, A_0(0) = - \frac{\eta}{1-\eta^2} + \frac{\eta}{(1-\eta^2)^{3/2} } \ln \frac{1+\sqrt{1-\eta^2}}{\eta},\ann
A_1(0) &=& -\eta (2-\pi\eta) + \frac{\eta(1-2\eta^2)}{\sqrt{1-\eta^2} } \ln \frac{1+\sqrt{1-\eta^2}}{\eta},\ann
A_2(0) &=& \eta (-\frac{7}{3}+ 4\eta^2) + \pi \eta^2 (\frac 32- 2\eta^2)  +\eta (1-4\eta^2)\sqrt{1-\eta^2} \ln \frac{1+\sqrt{1-\eta^2}}{\eta} ,\ann
A_3(0) &=& \eta ( -\frac{38}{15}+ 9\eta^2 - 6\eta^4 ) + \pi \eta^2 ( \frac{15}{8}- 5\eta^2 +3\eta^4)  + \eta (1-6\eta^2)(1-\eta^2)^{3/2} \ln \frac{1+\sqrt{1-\eta^2}}{\eta}\ann
B_0(0) &=& \frac{\eta^3}{1-\eta^2} + \frac{\pi \eta^2}{2} - \frac{\eta^3(2-\eta^2)}{(1-\eta^2)^{3/2} } \ln \frac{1+\sqrt{1-\eta^2}}{\eta}, B_1(0) = 3 \eta^3 + \frac{\pi \eta^2}{4} (1 - 6\eta^2) - \frac{\eta^3(2- 3\eta^2)}{\sqrt{1-\eta^2} } \ln \frac{1+\sqrt{1-\eta^2}}{\eta},\ann
B_2(0) &=& \frac{\eta^3}{3} (11-15 \eta^2) + \frac{\pi \eta^2}{16} (3 - 36\eta^2 + 40 \eta^4) - \eta^3(2- 5\eta^2)\sqrt{1-\eta^2} \ln \frac{1+\sqrt{1-\eta^2}}{\eta}.
\end{eqnarray}
\end{widetext}
Taking into account the above results, the TM asymptotic energy is found as 
\begin{eqnarray}
\mathcal{E}_{\TM}^{s,\as}(0) = \frac{Q_{\TM}^{s,\as} (\eta)}{R},
\end{eqnarray}
where
\begin{eqnarray}
Q_{\TM}^{s,\as}(\eta) &=& -\frac{\eta}{96\pi } (4(10-21\eta^2 ) - 3\pi \eta (9-14\eta^2 ) )\ann
&-& \frac{\eta^3}{8\pi} \frac{8-7\eta^2}{\sqrt{1-\eta^2}} \ln \frac{1+\sqrt{1-\eta^2}}{\eta}. 
\end{eqnarray}
\cite{Bulgac:2006:SCebDsoapaas}
\begin{acknowledgements}
NK is grateful for the financial support through the Fulbright Visiting Scholar Program and the hospitality of Department of Physics at the University of South Florida. NK was supported in part by the Russian Foundation for Basic Research Grant No. 13-02-00757-a. LMW acknowledges financial support from the Department of Energy under Contract No. DE-FG02-06ER46297. 
\end{acknowledgements}
\input{sigconst.bbl.tex}
\end{document}

%% file: sigconst.bbl.tex
%

%% file: sigconst.bbl
\begin{thebibliography}{52}%
\makeatletter
\providecommand \@ifxundefined [1]{%
 \@ifx{#1\undefined}
}%
\providecommand \@ifnum [1]{%
 \ifnum #1\expandafter \@firstoftwo
 \else \expandafter \@secondoftwo
 \fi
}%
\providecommand \@ifx [1]{%
 \ifx #1\expandafter \@firstoftwo
 \else \expandafter \@secondoftwo
 \fi
}%
\providecommand \natexlab [1]{#1}%
\providecommand \enquote  [1]{``#1''}%
\providecommand \bibnamefont  [1]{#1}%
\providecommand \bibfnamefont [1]{#1}%
\providecommand \citenamefont [1]{#1}%
\providecommand \href@noop [0]{\@secondoftwo}%
\providecommand \href [0]{\begingroup \@sanitize@url \@href}%
\providecommand \@href[1]{\@@startlink{#1}\@@href}%
\providecommand \@@href[1]{\endgroup#1\@@endlink}%
\providecommand \@sanitize@url [0]{\catcode `\\12\catcode `\$12\catcode
  `\&12\catcode `\#12\catcode `\^12\catcode `\_12\catcode `\%12\relax}%
\providecommand \@@startlink[1]{}%
\providecommand \@@endlink[0]{}%
\providecommand \url  [0]{\begingroup\@sanitize@url \@url }%
\providecommand \@url [1]{\endgroup\@href {#1}{\urlprefix }}%
\providecommand \urlprefix  [0]{URL }%
\providecommand \Eprint [0]{\href }%
\providecommand \doibase [0]{http://dx.doi.org/}%
\providecommand \selectlanguage [0]{\@gobble}%
\providecommand \bibinfo  [0]{\@secondoftwo}%
\providecommand \bibfield  [0]{\@secondoftwo}%
\providecommand \translation [1]{[#1]}%
\providecommand \BibitemOpen [0]{}%
\providecommand \bibitemStop [0]{}%
\providecommand \bibitemNoStop [0]{.\EOS\space}%
\providecommand \EOS [0]{\spacefactor3000\relax}%
\providecommand \BibitemShut  [1]{\csname bibitem#1\endcsname}%
\let\auto@bib@innerbib\@empty
\bibitem [{\citenamefont {Casimir}(1948)}]{Casimir:1948:Main}%
  \BibitemOpen
  \bibfield  {author} {\bibinfo {author} {\bibfnamefont {H.~B.~G.}\
  \bibnamefont {Casimir}},\ }\href@noop {} {\bibfield  {journal} {\bibinfo
  {journal} {Proc. K. Ned. Akad. Wet.}\ }\textbf {\bibinfo {volume} {51}},\
  \bibinfo {pages} {793} (\bibinfo {year} {1948})}\BibitemShut {NoStop}%
\bibitem [{\citenamefont {Casimir}\ and\ \citenamefont
  {Polder}(1948)}]{Casimir:1948:TIrLdWf}%
  \BibitemOpen
  \bibfield  {author} {\bibinfo {author} {\bibfnamefont {H.~B.~G.}\
  \bibnamefont {Casimir}}\ and\ \bibinfo {author} {\bibfnamefont
  {D.}~\bibnamefont {Polder}},\ }\href {\doibase 10.1103/PhysRev.73.360}
  {\bibfield  {journal} {\bibinfo  {journal} {Phys. Rev.}\ }\textbf {\bibinfo
  {volume} {73}},\ \bibinfo {pages} {360} (\bibinfo {year} {1948})}\BibitemShut
  {NoStop}%
\bibitem [{\citenamefont {Bordag}\ \emph {et~al.}(2001)\citenamefont {Bordag},
  \citenamefont {Mohideen},\ and\ \citenamefont
  {Mostepanenko}}]{Bordag:2001:NdCe}%
  \BibitemOpen
  \bibfield  {author} {\bibinfo {author} {\bibfnamefont {M.}~\bibnamefont
  {Bordag}}, \bibinfo {author} {\bibfnamefont {U.}~\bibnamefont {Mohideen}}, \
  and\ \bibinfo {author} {\bibfnamefont {V.~M.}\ \bibnamefont {Mostepanenko}},\
  }\href {\doibase 10.1016/S0370-1573(01)00015-1} {\bibfield  {journal}
  {\bibinfo  {journal} {Phys. Rept.}\ }\textbf {\bibinfo {volume} {353}},\
  \bibinfo {pages} {1} (\bibinfo {year} {2001})}\BibitemShut {NoStop}%
\bibitem [{\citenamefont {Milton}(2004)}]{Milton:2004:TCercap}%
  \BibitemOpen
  \bibfield  {author} {\bibinfo {author} {\bibfnamefont {K.~A.}\ \bibnamefont
  {Milton}},\ }\href {\doibase 10.1088/0305-4470/37/38/R01} {\bibfield
  {journal} {\bibinfo  {journal} {J. Phys. A: Math. Gen.}\ }\textbf {\bibinfo
  {volume} {37}},\ \bibinfo {pages} {R209} (\bibinfo {year}
  {2004})}\BibitemShut {NoStop}%
\bibitem [{\citenamefont {Buhmann}\ and\ \citenamefont
  {Welsch}(2007)}]{Buhmann:2007:Dfimqe}%
  \BibitemOpen
  \bibfield  {author} {\bibinfo {author} {\bibfnamefont {S.~Y.}\ \bibnamefont
  {Buhmann}}\ and\ \bibinfo {author} {\bibfnamefont {D.-G.}\ \bibnamefont
  {Welsch}},\ }\href {\doibase 10.1016/j.pquantelec.2007.03.001} {\bibfield
  {journal} {\bibinfo  {journal} {Prog. Quant. Electron.}\ }\textbf {\bibinfo
  {volume} {31}},\ \bibinfo {pages} {51} (\bibinfo {year} {2007})}\BibitemShut
  {NoStop}%
\bibitem [{\citenamefont {Klimchitskaya}\ \emph {et~al.}(2009)\citenamefont
  {Klimchitskaya}, \citenamefont {Mohideen},\ and\ \citenamefont
  {Mostepanenko}}]{Klimchitskaya:2009:TCfbrmet}%
  \BibitemOpen
  \bibfield  {author} {\bibinfo {author} {\bibfnamefont {G.~L.}\ \bibnamefont
  {Klimchitskaya}}, \bibinfo {author} {\bibfnamefont {U.}~\bibnamefont
  {Mohideen}}, \ and\ \bibinfo {author} {\bibfnamefont {V.~M.}\ \bibnamefont
  {Mostepanenko}},\ }\href {\doibase 10.1103/RevModPhys.81.1827} {\bibfield
  {journal} {\bibinfo  {journal} {Rev. Mod. Phys.}\ }\textbf {\bibinfo {volume}
  {81}},\ \bibinfo {pages} {1827} (\bibinfo {year} {2009})}\BibitemShut
  {NoStop}%
\bibitem [{\citenamefont {Milton}(2001)}]{Milton:2001:CEPMZE}%
  \BibitemOpen
  \bibfield  {author} {\bibinfo {author} {\bibfnamefont {K.~A.}\ \bibnamefont
  {Milton}},\ }\href@noop {} {\emph {\bibinfo {title} {Casimir Effect: Physical
  Manifestation of Zero-Point Energy.}}}\ (\bibinfo  {publisher} {World
  Scientific Publishing Co.},\ \bibinfo {year} {2001})\ p.\ \bibinfo {pages}
  {301}\BibitemShut {NoStop}%
\bibitem [{\citenamefont {Bordag}\ \emph
  {et~al.}(2009{\natexlab{a}})\citenamefont {Bordag}, \citenamefont
  {Klimchitskaya}, \citenamefont {Mohideen},\ and\ \citenamefont
  {Mostepanenko}}]{Bordag:2009:ACE}%
  \BibitemOpen
  \bibfield  {author} {\bibinfo {author} {\bibfnamefont {M.}~\bibnamefont
  {Bordag}}, \bibinfo {author} {\bibfnamefont {G.}~\bibnamefont
  {Klimchitskaya}}, \bibinfo {author} {\bibfnamefont {U.}~\bibnamefont
  {Mohideen}}, \ and\ \bibinfo {author} {\bibfnamefont {V.}~\bibnamefont
  {Mostepanenko}},\ }\href@noop {} {\emph {\bibinfo {title} {Advances in the
  Casimir Effect}}}\ (\bibinfo  {publisher} {Oxford University Press, Oxford},\
  \bibinfo {year} {2009})\ p.\ \bibinfo {pages} {749}\BibitemShut {NoStop}%
\bibitem [{\citenamefont {Banishev}\ \emph
  {et~al.}(2013{\natexlab{a}})\citenamefont {Banishev}, \citenamefont {Wagner},
  \citenamefont {Emig}, \citenamefont {Zandi},\ and\ \citenamefont
  {Mohideen}}]{Banishev:2013:DoACFbC}%
  \BibitemOpen
  \bibfield  {author} {\bibinfo {author} {\bibfnamefont {A.~A.}\ \bibnamefont
  {Banishev}}, \bibinfo {author} {\bibfnamefont {J.}~\bibnamefont {Wagner}},
  \bibinfo {author} {\bibfnamefont {T.}~\bibnamefont {Emig}}, \bibinfo {author}
  {\bibfnamefont {R.}~\bibnamefont {Zandi}}, \ and\ \bibinfo {author}
  {\bibfnamefont {U.}~\bibnamefont {Mohideen}},\ }\href {\doibase
  10.1103/PhysRevLett.110.250403} {\bibfield  {journal} {\bibinfo  {journal}
  {Phys. Rev. Lett.}\ }\textbf {\bibinfo {volume} {110}},\ \bibinfo {pages}
  {250403} (\bibinfo {year} {2013}{\natexlab{a}})}\BibitemShut {NoStop}%
\bibitem [{\citenamefont {Emig}\ \emph {et~al.}(2006)\citenamefont {Emig},
  \citenamefont {Jaffe}, \citenamefont {Kardar},\ and\ \citenamefont
  {Scardicchio}}]{Emig:2006:CIbaPaaC}%
  \BibitemOpen
  \bibfield  {author} {\bibinfo {author} {\bibfnamefont {T.}~\bibnamefont
  {Emig}}, \bibinfo {author} {\bibfnamefont {R.~L.}\ \bibnamefont {Jaffe}},
  \bibinfo {author} {\bibfnamefont {M.}~\bibnamefont {Kardar}}, \ and\ \bibinfo
  {author} {\bibfnamefont {A.}~\bibnamefont {Scardicchio}},\ }\href {\doibase
  10.1103/PhysRevLett.96.080403} {\bibfield  {journal} {\bibinfo  {journal}
  {Phys. Rev. Lett.}\ }\textbf {\bibinfo {volume} {96}},\ \bibinfo {pages}
  {080403} (\bibinfo {year} {2006})}\BibitemShut {NoStop}%
\bibitem [{\citenamefont {Bulgac}\ \emph
  {et~al.}(2006{\natexlab{a}})\citenamefont {Bulgac}, \citenamefont
  {Magierski},\ and\ \citenamefont {Wirzba}}]{Bulgac:2006:SCebDsops}%
  \BibitemOpen
  \bibfield  {author} {\bibinfo {author} {\bibfnamefont {A.}~\bibnamefont
  {Bulgac}}, \bibinfo {author} {\bibfnamefont {P.}~\bibnamefont {Magierski}}, \
  and\ \bibinfo {author} {\bibfnamefont {A.}~\bibnamefont {Wirzba}},\ }\href
  {\doibase 10.1103/PhysRevD.73.025007} {\bibfield  {journal} {\bibinfo
  {journal} {Phys. Rev. D}\ }\textbf {\bibinfo {volume} {73}},\ \bibinfo
  {pages} {025007} (\bibinfo {year} {2006}{\natexlab{a}})}\BibitemShut
  {NoStop}%
\bibitem [{\citenamefont {Novoselov}\ \emph {et~al.}(2004)\citenamefont
  {Novoselov}, \citenamefont {Geim}, \citenamefont {Morozov}, \citenamefont
  {Jiang}, \citenamefont {Zhang}, \citenamefont {Dubonos}, \citenamefont
  {Grigorieva},\ and\ \citenamefont {Firsov}}]{Novoselov:2004:EFEiATCF}%
  \BibitemOpen
  \bibfield  {author} {\bibinfo {author} {\bibfnamefont {K.~S.}\ \bibnamefont
  {Novoselov}}, \bibinfo {author} {\bibfnamefont {A.~K.}\ \bibnamefont {Geim}},
  \bibinfo {author} {\bibfnamefont {S.~V.}\ \bibnamefont {Morozov}}, \bibinfo
  {author} {\bibfnamefont {D.}~\bibnamefont {Jiang}}, \bibinfo {author}
  {\bibfnamefont {Y.}~\bibnamefont {Zhang}}, \bibinfo {author} {\bibfnamefont
  {S.~V.}\ \bibnamefont {Dubonos}}, \bibinfo {author} {\bibfnamefont {I.~V.}\
  \bibnamefont {Grigorieva}}, \ and\ \bibinfo {author} {\bibfnamefont {A.~A.}\
  \bibnamefont {Firsov}},\ }\href {\doibase 10.1126/science.1102896} {\bibfield
   {journal} {\bibinfo  {journal} {Science}\ }\textbf {\bibinfo {volume}
  {306}},\ \bibinfo {pages} {666} (\bibinfo {year} {2004})}\BibitemShut
  {NoStop}%
\bibitem [{\citenamefont {Dobson}\ \emph {et~al.}(2006)\citenamefont {Dobson},
  \citenamefont {White},\ and\ \citenamefont {Rubio}}]{Dobson:2006:ADIABvdWEF}%
  \BibitemOpen
  \bibfield  {author} {\bibinfo {author} {\bibfnamefont {J.~F.}\ \bibnamefont
  {Dobson}}, \bibinfo {author} {\bibfnamefont {A.}~\bibnamefont {White}}, \
  and\ \bibinfo {author} {\bibfnamefont {A.}~\bibnamefont {Rubio}},\
  }\href@noop {} {\bibfield  {journal} {\bibinfo  {journal} {Phys. Rev. Lett.}\
  }\textbf {\bibinfo {volume} {96}},\ \bibinfo {pages} {073201} (\bibinfo
  {year} {2006})}\BibitemShut {NoStop}%
\bibitem [{\citenamefont {G\'omez-Santos}(2009)}]{Gomez-Santos:2009:TvdWibgl}%
  \BibitemOpen
  \bibfield  {author} {\bibinfo {author} {\bibfnamefont {G.}~\bibnamefont
  {G\'omez-Santos}},\ }\href {\doibase 10.1103/PhysRevB.80.245424} {\bibfield
  {journal} {\bibinfo  {journal} {Phys. Rev. B}\ }\textbf {\bibinfo {volume}
  {80}},\ \bibinfo {pages} {245424} (\bibinfo {year} {2009})}\BibitemShut
  {NoStop}%
\bibitem [{\citenamefont {Drosdoff}\ and\ \citenamefont
  {Woods}(2010)}]{Drosdoff:2010:Cfgs}%
  \BibitemOpen
  \bibfield  {author} {\bibinfo {author} {\bibfnamefont {D.}~\bibnamefont
  {Drosdoff}}\ and\ \bibinfo {author} {\bibfnamefont {L.~M.}\ \bibnamefont
  {Woods}},\ }\href {\doibase 10.1103/PhysRevB.82.155459} {\bibfield  {journal}
  {\bibinfo  {journal} {Phys. Rev. B}\ }\textbf {\bibinfo {volume} {82}},\
  \bibinfo {pages} {155459} (\bibinfo {year} {2010})}\BibitemShut {NoStop}%
\bibitem [{\citenamefont {Drosdoff}\ \emph {et~al.}(2012)\citenamefont
  {Drosdoff}, \citenamefont {Phan}, \citenamefont {Woods}, \citenamefont
  {Bondarev},\ and\ \citenamefont {Dobson}}]{Drosdoff:2012:EosdotCfbgs}%
  \BibitemOpen
  \bibfield  {author} {\bibinfo {author} {\bibfnamefont {D.}~\bibnamefont
  {Drosdoff}}, \bibinfo {author} {\bibfnamefont {A.}~\bibnamefont {Phan}},
  \bibinfo {author} {\bibfnamefont {L.}~\bibnamefont {Woods}}, \bibinfo
  {author} {\bibfnamefont {I.}~\bibnamefont {Bondarev}}, \ and\ \bibinfo
  {author} {\bibfnamefont {J.}~\bibnamefont {Dobson}},\ }\href {\doibase
  10.1140/epjb/e2012-30741-6} {\bibfield  {journal} {\bibinfo  {journal} {Eur.
  Phys J. B}\ }\textbf {\bibinfo {volume} {85}},\ \bibinfo {pages} {1}
  (\bibinfo {year} {2012})}\BibitemShut {NoStop}%
\bibitem [{\citenamefont {Klimchitskaya}\ and\ \citenamefont
  {Mostepanenko}(2013)}]{Klimchitskaya:2013:vdWaCibtgs}%
  \BibitemOpen
  \bibfield  {author} {\bibinfo {author} {\bibfnamefont {G.~L.}\ \bibnamefont
  {Klimchitskaya}}\ and\ \bibinfo {author} {\bibfnamefont {V.~M.}\ \bibnamefont
  {Mostepanenko}},\ }\href {\doibase 10.1103/PhysRevB.87.075439} {\bibfield
  {journal} {\bibinfo  {journal} {Phys. Rev. B}\ }\textbf {\bibinfo {volume}
  {87}},\ \bibinfo {pages} {075439} (\bibinfo {year} {2013})}\BibitemShut
  {NoStop}%
\bibitem [{\citenamefont {Banishev}\ \emph
  {et~al.}(2013{\natexlab{b}})\citenamefont {Banishev}, \citenamefont {Wen},
  \citenamefont {Xu}, \citenamefont {Kawakami}, \citenamefont {Klimchitskaya},
  \citenamefont {Mostepanenko},\ and\ \citenamefont
  {Mohideen}}]{Banishev:2013:MtCfgfgoaSIOS}%
  \BibitemOpen
  \bibfield  {author} {\bibinfo {author} {\bibfnamefont {A.~A.}\ \bibnamefont
  {Banishev}}, \bibinfo {author} {\bibfnamefont {H.}~\bibnamefont {Wen}},
  \bibinfo {author} {\bibfnamefont {J.}~\bibnamefont {Xu}}, \bibinfo {author}
  {\bibfnamefont {R.~K.}\ \bibnamefont {Kawakami}}, \bibinfo {author}
  {\bibfnamefont {G.~L.}\ \bibnamefont {Klimchitskaya}}, \bibinfo {author}
  {\bibfnamefont {V.~M.}\ \bibnamefont {Mostepanenko}}, \ and\ \bibinfo
  {author} {\bibfnamefont {U.}~\bibnamefont {Mohideen}},\ }\href {\doibase
  10.1103/PhysRevB.87.205433} {\bibfield  {journal} {\bibinfo  {journal} {Phys.
  Rev. B}\ }\textbf {\bibinfo {volume} {87}},\ \bibinfo {pages} {205433}
  (\bibinfo {year} {2013}{\natexlab{b}})}\BibitemShut {NoStop}%
\bibitem [{\citenamefont {Bordag}\ \emph {et~al.}(2006)\citenamefont {Bordag},
  \citenamefont {Geyer}, \citenamefont {Klimchitskaya},\ and\ \citenamefont
  {Mostepanenko}}]{Bordag:2006:LffgascnvdWaCi}%
  \BibitemOpen
  \bibfield  {author} {\bibinfo {author} {\bibfnamefont {M.}~\bibnamefont
  {Bordag}}, \bibinfo {author} {\bibfnamefont {B.}~\bibnamefont {Geyer}},
  \bibinfo {author} {\bibfnamefont {G.~L.}\ \bibnamefont {Klimchitskaya}}, \
  and\ \bibinfo {author} {\bibfnamefont {V.~M.}\ \bibnamefont {Mostepanenko}},\
  }\href {\doibase 10.1103/PhysRevB.74.205431} {\bibfield  {journal} {\bibinfo
  {journal} {Phys. Rev. B}\ }\textbf {\bibinfo {volume} {74}},\ \bibinfo
  {pages} {205431} (\bibinfo {year} {2006})}\BibitemShut {NoStop}%
\bibitem [{\citenamefont {Bordag}\ \emph
  {et~al.}(2009{\natexlab{b}})\citenamefont {Bordag}, \citenamefont
  {Fialkovsky}, \citenamefont {Gitman},\ and\ \citenamefont
  {Vassilevich}}]{Bordag:2009:CibapcagdbtDm}%
  \BibitemOpen
  \bibfield  {author} {\bibinfo {author} {\bibfnamefont {M.}~\bibnamefont
  {Bordag}}, \bibinfo {author} {\bibfnamefont {I.~V.}\ \bibnamefont
  {Fialkovsky}}, \bibinfo {author} {\bibfnamefont {D.~M.}\ \bibnamefont
  {Gitman}}, \ and\ \bibinfo {author} {\bibfnamefont {D.~V.}\ \bibnamefont
  {Vassilevich}},\ }\href {\doibase 10.1103/PhysRevB.80.245406} {\bibfield
  {journal} {\bibinfo  {journal} {Phys. Rev. B}\ }\textbf {\bibinfo {volume}
  {80}},\ \bibinfo {pages} {245406} (\bibinfo {year}
  {2009}{\natexlab{b}})}\BibitemShut {NoStop}%
\bibitem [{\citenamefont {Drosdoff}\ and\ \citenamefont
  {Woods}(2011)}]{Drosdoff:2011:Cibgsam}%
  \BibitemOpen
  \bibfield  {author} {\bibinfo {author} {\bibfnamefont {D.}~\bibnamefont
  {Drosdoff}}\ and\ \bibinfo {author} {\bibfnamefont {L.~M.}\ \bibnamefont
  {Woods}},\ }\href {\doibase 10.1103/PhysRevA.84.062501} {\bibfield  {journal}
  {\bibinfo  {journal} {Phys. Rev. A}\ }\textbf {\bibinfo {volume} {84}},\
  \bibinfo {pages} {062501} (\bibinfo {year} {2011})}\BibitemShut {NoStop}%
\bibitem [{\citenamefont {Phan}\ \emph {et~al.}(2012)\citenamefont {Phan},
  \citenamefont {Viet}, \citenamefont {Poklonski}, \citenamefont {Woods},\ and\
  \citenamefont {Le}}]{Phan:2012:Ioagswafmp}%
  \BibitemOpen
  \bibfield  {author} {\bibinfo {author} {\bibfnamefont {A.~D.}\ \bibnamefont
  {Phan}}, \bibinfo {author} {\bibfnamefont {N.~A.}\ \bibnamefont {Viet}},
  \bibinfo {author} {\bibfnamefont {N.~A.}\ \bibnamefont {Poklonski}}, \bibinfo
  {author} {\bibfnamefont {L.~M.}\ \bibnamefont {Woods}}, \ and\ \bibinfo
  {author} {\bibfnamefont {C.~H.}\ \bibnamefont {Le}},\ }\href {\doibase
  10.1103/PhysRevB.86.155419} {\bibfield  {journal} {\bibinfo  {journal} {Phys.
  Rev. B}\ }\textbf {\bibinfo {volume} {86}},\ \bibinfo {pages} {155419}
  (\bibinfo {year} {2012})}\BibitemShut {NoStop}%
\bibitem [{\citenamefont {Bordag}\ \emph {et~al.}(2012)\citenamefont {Bordag},
  \citenamefont {Klimchitskaya},\ and\ \citenamefont
  {Mostepanenko}}]{Bordag:2012:TCeitiogwdam}%
  \BibitemOpen
  \bibfield  {author} {\bibinfo {author} {\bibfnamefont {M.}~\bibnamefont
  {Bordag}}, \bibinfo {author} {\bibfnamefont {G.~L.}\ \bibnamefont
  {Klimchitskaya}}, \ and\ \bibinfo {author} {\bibfnamefont {V.~M.}\
  \bibnamefont {Mostepanenko}},\ }\href {\doibase 10.1103/PhysRevB.86.165429}
  {\bibfield  {journal} {\bibinfo  {journal} {Phys. Rev. B}\ }\textbf {\bibinfo
  {volume} {86}},\ \bibinfo {pages} {165429} (\bibinfo {year}
  {2012})}\BibitemShut {NoStop}%
\bibitem [{\citenamefont {Di\~{n}o}\ \emph {et~al.}(2004)\citenamefont
  {Di\~{n}o}, \citenamefont {Nakanishi}, \citenamefont {Kasai}, \citenamefont
  {Sugimoto},\ and\ \citenamefont {Kondo}}]{Dino:2004:DAatZEoG}%
  \BibitemOpen
  \bibfield  {author} {\bibinfo {author} {\bibfnamefont {W.~A.}\ \bibnamefont
  {Di\~{n}o}}, \bibinfo {author} {\bibfnamefont {H.}~\bibnamefont {Nakanishi}},
  \bibinfo {author} {\bibfnamefont {H.}~\bibnamefont {Kasai}}, \bibinfo
  {author} {\bibfnamefont {T.}~\bibnamefont {Sugimoto}}, \ and\ \bibinfo
  {author} {\bibfnamefont {T.}~\bibnamefont {Kondo}},\ }\href {\doibase
  10.1380/ejssnt.2004.77} {\bibfield  {journal} {\bibinfo  {journal} {e-J.
  Surf. Sci. Nanotech.}\ }\textbf {\bibinfo {volume} {2}},\ \bibinfo {pages}
  {77} (\bibinfo {year} {2004})}\BibitemShut {NoStop}%
\bibitem [{\citenamefont {Bondarev}\ and\ \citenamefont
  {Lambin}(2004)}]{Bondarev:2004:vdWeusacdcn}%
  \BibitemOpen
  \bibfield  {author} {\bibinfo {author} {\bibfnamefont {I.}~\bibnamefont
  {Bondarev}}\ and\ \bibinfo {author} {\bibfnamefont {P.}~\bibnamefont
  {Lambin}},\ }\href {\doibase 10.1016/j.ssc.2004.07.039} {\bibfield  {journal}
  {\bibinfo  {journal} {Solid State Commun.}\ }\textbf {\bibinfo {volume}
  {132}},\ \bibinfo {pages} {203} (\bibinfo {year} {2004})}\BibitemShut
  {NoStop}%
\bibitem [{\citenamefont {Bondarev}\ and\ \citenamefont
  {Lambin}(2005)}]{Bondarev:2005:vdWciadcn}%
  \BibitemOpen
  \bibfield  {author} {\bibinfo {author} {\bibfnamefont {I.~V.}\ \bibnamefont
  {Bondarev}}\ and\ \bibinfo {author} {\bibfnamefont {P.}~\bibnamefont
  {Lambin}},\ }\href {\doibase 10.1103/PhysRevB.72.035451} {\bibfield
  {journal} {\bibinfo  {journal} {Phys. Rev. B}\ }\textbf {\bibinfo {volume}
  {72}},\ \bibinfo {pages} {035451} (\bibinfo {year} {2005})}\BibitemShut
  {NoStop}%
\bibitem [{\citenamefont {Blagov}\ \emph {et~al.}(2007)\citenamefont {Blagov},
  \citenamefont {Klimchitskaya},\ and\ \citenamefont
  {Mostepanenko}}]{Blagov:2007:vdWibamaascn}%
  \BibitemOpen
  \bibfield  {author} {\bibinfo {author} {\bibfnamefont {E.~V.}\ \bibnamefont
  {Blagov}}, \bibinfo {author} {\bibfnamefont {G.~L.}\ \bibnamefont
  {Klimchitskaya}}, \ and\ \bibinfo {author} {\bibfnamefont {V.~M.}\
  \bibnamefont {Mostepanenko}},\ }\href {\doibase 10.1103/PhysRevB.75.235413}
  {\bibfield  {journal} {\bibinfo  {journal} {Phys. Rev. B}\ }\textbf {\bibinfo
  {volume} {75}},\ \bibinfo {pages} {235413} (\bibinfo {year}
  {2007})}\BibitemShut {NoStop}%
\bibitem [{\citenamefont {Churkin}\ \emph {et~al.}(2010)\citenamefont
  {Churkin}, \citenamefont {Fedortsov}, \citenamefont {Klimchitskaya},\ and\
  \citenamefont {Yurova}}]{Churkin:2010:CohaDmodibgaHHoNa}%
  \BibitemOpen
  \bibfield  {author} {\bibinfo {author} {\bibfnamefont {Y.~V.}\ \bibnamefont
  {Churkin}}, \bibinfo {author} {\bibfnamefont {A.~B.}\ \bibnamefont
  {Fedortsov}}, \bibinfo {author} {\bibfnamefont {G.~L.}\ \bibnamefont
  {Klimchitskaya}}, \ and\ \bibinfo {author} {\bibfnamefont {V.~A.}\
  \bibnamefont {Yurova}},\ }\href {\doibase 10.1103/PhysRevB.82.165433}
  {\bibfield  {journal} {\bibinfo  {journal} {Phys. Rev. B}\ }\textbf {\bibinfo
  {volume} {82}},\ \bibinfo {pages} {165433} (\bibinfo {year}
  {2010})}\BibitemShut {NoStop}%
\bibitem [{\citenamefont {Chaichian}\ \emph {et~al.}(2012)\citenamefont
  {Chaichian}, \citenamefont {Klimchitskaya}, \citenamefont {Mostepanenko},\
  and\ \citenamefont {Tureanu}}]{Chaichian:2012:TCiodawg}%
  \BibitemOpen
  \bibfield  {author} {\bibinfo {author} {\bibfnamefont {M.}~\bibnamefont
  {Chaichian}}, \bibinfo {author} {\bibfnamefont {G.~L.}\ \bibnamefont
  {Klimchitskaya}}, \bibinfo {author} {\bibfnamefont {V.~M.}\ \bibnamefont
  {Mostepanenko}}, \ and\ \bibinfo {author} {\bibfnamefont {A.}~\bibnamefont
  {Tureanu}},\ }\href {\doibase 10.1103/PhysRevA.86.012515} {\bibfield
  {journal} {\bibinfo  {journal} {Phys. Rev. A}\ }\textbf {\bibinfo {volume}
  {86}},\ \bibinfo {pages} {012515} (\bibinfo {year} {2012})}\BibitemShut
  {NoStop}%
\bibitem [{\citenamefont {Barton}(2004)}]{Barton:2004:Cesps}%
  \BibitemOpen
  \bibfield  {author} {\bibinfo {author} {\bibfnamefont {G.}~\bibnamefont
  {Barton}},\ }\href {\doibase 10.1088/0305-4470/37/3/032} {\bibfield
  {journal} {\bibinfo  {journal} {J. Phys. A: Math. Gen.}\ }\textbf {\bibinfo
  {volume} {37}},\ \bibinfo {pages} {1011} (\bibinfo {year}
  {2004})}\BibitemShut {NoStop}%
\bibitem [{\citenamefont {Barton}(2005)}]{Barton:2005:CefpsIE}%
  \BibitemOpen
  \bibfield  {author} {\bibinfo {author} {\bibfnamefont {G.}~\bibnamefont
  {Barton}},\ }\href {\doibase 10.1088/0305-4470/38/13/013} {\bibfield
  {journal} {\bibinfo  {journal} {J. Phys. A: Math. Gen.}\ }\textbf {\bibinfo
  {volume} {38}},\ \bibinfo {pages} {2997} (\bibinfo {year}
  {2005})}\BibitemShut {NoStop}%
\bibitem [{\citenamefont {Bordag}(2006)}]{Bordag:2006:TCeftpsatrotsp}%
  \BibitemOpen
  \bibfield  {author} {\bibinfo {author} {\bibfnamefont {M.}~\bibnamefont
  {Bordag}},\ }\href {\doibase 10.1088/0305-4470/39/21/S08} {\bibfield
  {journal} {\bibinfo  {journal} {J. Phys. A: Math. Gen.}\ }\textbf {\bibinfo
  {volume} {39}},\ \bibinfo {pages} {6173} (\bibinfo {year}
  {2006})}\BibitemShut {NoStop}%
\bibitem [{\citenamefont {Bordag}\ \emph {et~al.}(1992)\citenamefont {Bordag},
  \citenamefont {Hennig},\ and\ \citenamefont
  {Robaschik}}]{Bordag:1992:Veiqftwepcop}%
  \BibitemOpen
  \bibfield  {author} {\bibinfo {author} {\bibfnamefont {M.}~\bibnamefont
  {Bordag}}, \bibinfo {author} {\bibfnamefont {D.}~\bibnamefont {Hennig}}, \
  and\ \bibinfo {author} {\bibfnamefont {D.}~\bibnamefont {Robaschik}},\ }\href
  {\doibase 10.1088/0305-4470/25/16/023} {\bibfield  {journal} {\bibinfo
  {journal} {J. Phys. A: Math. Gen.}\ }\textbf {\bibinfo {volume} {25}},\
  \bibinfo {pages} {4483} (\bibinfo {year} {1992})}\BibitemShut {NoStop}%
\bibitem [{\citenamefont {Bordag}(2007)}]{Bordag:2007:Ioacwatps}%
  \BibitemOpen
  \bibfield  {author} {\bibinfo {author} {\bibfnamefont {M.}~\bibnamefont
  {Bordag}},\ }\href {\doibase 10.1103/PhysRevD.76.065011} {\bibfield
  {journal} {\bibinfo  {journal} {Phys. Rev. D}\ }\textbf {\bibinfo {volume}
  {76}},\ \bibinfo {pages} {065011} (\bibinfo {year} {2007})}\BibitemShut
  {NoStop}%
\bibitem [{\citenamefont {Fetter}(1973)}]{Fetter:1973:EoalegISl}%
  \BibitemOpen
  \bibfield  {author} {\bibinfo {author} {\bibfnamefont {A.~L.}\ \bibnamefont
  {Fetter}},\ }\href {\doibase http://dx.doi.org/10.1016/0003-4916(73)90161-9}
  {\bibfield  {journal} {\bibinfo  {journal} {Ann. Phys. New York}\ }\textbf
  {\bibinfo {volume} {81}},\ \bibinfo {pages} {367 } (\bibinfo {year}
  {1973})}\BibitemShut {NoStop}%
\bibitem [{\citenamefont {Fetter}(1974)}]{Fetter:1974:EoalegIPa}%
  \BibitemOpen
  \bibfield  {author} {\bibinfo {author} {\bibfnamefont {A.~L.}\ \bibnamefont
  {Fetter}},\ }\href {\doibase http://dx.doi.org/10.1016/0003-4916(74)90397-2}
  {\bibfield  {journal} {\bibinfo  {journal} {Ann. Phys. New York}\ }\textbf
  {\bibinfo {volume} {88}},\ \bibinfo {pages} {1 } (\bibinfo {year}
  {1974})}\BibitemShut {NoStop}%
\bibitem [{\citenamefont {Gusynin}\ \emph {et~al.}(2007)\citenamefont
  {Gusynin}, \citenamefont {Sharapov},\ and\ \citenamefont
  {Carbotte}}]{Gusynin:2007:Mcig}%
  \BibitemOpen
  \bibfield  {author} {\bibinfo {author} {\bibfnamefont {V.~P.}\ \bibnamefont
  {Gusynin}}, \bibinfo {author} {\bibfnamefont {S.~G.}\ \bibnamefont
  {Sharapov}}, \ and\ \bibinfo {author} {\bibfnamefont {J.~P.}\ \bibnamefont
  {Carbotte}},\ }\href {\doibase 10.1088/0953-8984/19/2/026222} {\bibfield
  {journal} {\bibinfo  {journal} {J. Phys.: Condens. Matter}\ }\textbf
  {\bibinfo {volume} {19}},\ \bibinfo {pages} {026222} (\bibinfo {year}
  {2007})}\BibitemShut {NoStop}%
\bibitem [{\citenamefont {Falkovsky}\ and\ \citenamefont
  {Varlamov}(2007)}]{Falkovsky:2007:Sdogc}%
  \BibitemOpen
  \bibfield  {author} {\bibinfo {author} {\bibfnamefont {L.~A.}\ \bibnamefont
  {Falkovsky}}\ and\ \bibinfo {author} {\bibfnamefont {A.~A.}\ \bibnamefont
  {Varlamov}},\ }\href {\doibase 10.1140/epjb/e2007-00142-3} {\bibfield
  {journal} {\bibinfo  {journal} {Eur. Phys J. B}\ }\textbf {\bibinfo {volume}
  {56}},\ \bibinfo {pages} {281} (\bibinfo {year} {2007})}\BibitemShut
  {NoStop}%
\bibitem [{\citenamefont {Nair}\ \emph {et~al.}(2008)\citenamefont {Nair},
  \citenamefont {Blake}, \citenamefont {Grigorenko}, \citenamefont {Novoselov},
  \citenamefont {Booth}, \citenamefont {Stauber}, \citenamefont {Peres},\ and\
  \citenamefont {Geim}}]{Nair:2008:FSCDVToG}%
  \BibitemOpen
  \bibfield  {author} {\bibinfo {author} {\bibfnamefont {R.~R.}\ \bibnamefont
  {Nair}}, \bibinfo {author} {\bibfnamefont {P.}~\bibnamefont {Blake}},
  \bibinfo {author} {\bibfnamefont {A.~N.}\ \bibnamefont {Grigorenko}},
  \bibinfo {author} {\bibfnamefont {K.~S.}\ \bibnamefont {Novoselov}}, \bibinfo
  {author} {\bibfnamefont {T.~J.}\ \bibnamefont {Booth}}, \bibinfo {author}
  {\bibfnamefont {T.}~\bibnamefont {Stauber}}, \bibinfo {author} {\bibfnamefont
  {N.~M.~R.}\ \bibnamefont {Peres}}, \ and\ \bibinfo {author} {\bibfnamefont
  {A.~K.}\ \bibnamefont {Geim}},\ }\href {\doibase 10.1126/science.1156965}
  {\bibfield  {journal} {\bibinfo  {journal} {Science}\ }\textbf {\bibinfo
  {volume} {320}},\ \bibinfo {pages} {1308} (\bibinfo {year}
  {2008})}\BibitemShut {NoStop}%
\bibitem [{\citenamefont {Bordag}\ \emph {et~al.}(1997)\citenamefont {Bordag},
  \citenamefont {Elizalde}, \citenamefont {Kirsten},\ and\ \citenamefont
  {Leseduarte}}]{Bordag:1997:Cefmfitb}%
  \BibitemOpen
  \bibfield  {author} {\bibinfo {author} {\bibfnamefont {M.}~\bibnamefont
  {Bordag}}, \bibinfo {author} {\bibfnamefont {E.}~\bibnamefont {Elizalde}},
  \bibinfo {author} {\bibfnamefont {K.}~\bibnamefont {Kirsten}}, \ and\
  \bibinfo {author} {\bibfnamefont {S.}~\bibnamefont {Leseduarte}},\ }\href
  {\doibase 10.1103/PhysRevD.56.4896} {\bibfield  {journal} {\bibinfo
  {journal} {Phys. Rev. D}\ }\textbf {\bibinfo {volume} {56}},\ \bibinfo
  {pages} {4896} (\bibinfo {year} {1997})}\BibitemShut {NoStop}%
\bibitem [{\citenamefont {Barash}\ and\ \citenamefont
  {Ginzburg}(1975)}]{Barash:1975:EfimamVfbt}%
  \BibitemOpen
  \bibfield  {author} {\bibinfo {author} {\bibfnamefont {Y.~S.}\ \bibnamefont
  {Barash}}\ and\ \bibinfo {author} {\bibfnamefont {V.~L.}\ \bibnamefont
  {Ginzburg}},\ }\href {\doibase 10.1070/PU1975v018n05ABEH001958} {\bibfield
  {journal} {\bibinfo  {journal} {Physics-Uspekhi}\ }\textbf {\bibinfo {volume}
  {18}},\ \bibinfo {pages} {305} (\bibinfo {year} {1975})}\BibitemShut
  {NoStop}%
\bibitem [{\citenamefont {Nesterenko}\ and\ \citenamefont
  {Pirozhenko}(2012)}]{Nesterenko:2012:Lfbassm}%
  \BibitemOpen
  \bibfield  {author} {\bibinfo {author} {\bibfnamefont {V.~V.}\ \bibnamefont
  {Nesterenko}}\ and\ \bibinfo {author} {\bibfnamefont {I.~G.}\ \bibnamefont
  {Pirozhenko}},\ }\href {\doibase 10.1103/PhysRevA.86.052503} {\bibfield
  {journal} {\bibinfo  {journal} {Phys. Rev. A}\ }\textbf {\bibinfo {volume}
  {86}},\ \bibinfo {pages} {052503} (\bibinfo {year} {2012})}\BibitemShut
  {NoStop}%
\bibitem [{\citenamefont {Parashar}\ \emph {et~al.}(2012)\citenamefont
  {Parashar}, \citenamefont {Milton}, \citenamefont {Shajesh},\ and\
  \citenamefont {Schaden}}]{Parashar:2012:EsdpCiebpitp}%
  \BibitemOpen
  \bibfield  {author} {\bibinfo {author} {\bibfnamefont {P.}~\bibnamefont
  {Parashar}}, \bibinfo {author} {\bibfnamefont {K.~A.}\ \bibnamefont
  {Milton}}, \bibinfo {author} {\bibfnamefont {K.~V.}\ \bibnamefont {Shajesh}},
  \ and\ \bibinfo {author} {\bibfnamefont {M.}~\bibnamefont {Schaden}},\ }\href
  {\doibase 10.1103/PhysRevD.86.085021} {\bibfield  {journal} {\bibinfo
  {journal} {Phys. Rev. D}\ }\textbf {\bibinfo {volume} {86}},\ \bibinfo
  {pages} {085021} (\bibinfo {year} {2012})}\BibitemShut {NoStop}%
\bibitem [{\citenamefont {Nesterenko}\ and\ \citenamefont
  {Pirozhenko}(1998)}]{Nesterenko:1998:SmfctCefas}%
  \BibitemOpen
  \bibfield  {author} {\bibinfo {author} {\bibfnamefont {V.~V.}\ \bibnamefont
  {Nesterenko}}\ and\ \bibinfo {author} {\bibfnamefont {I.~G.}\ \bibnamefont
  {Pirozhenko}},\ }\href {\doibase 10.1103/PhysRevD.57.1284} {\bibfield
  {journal} {\bibinfo  {journal} {Phys. Rev. D}\ }\textbf {\bibinfo {volume}
  {57}},\ \bibinfo {pages} {1284} (\bibinfo {year} {1998})}\BibitemShut
  {NoStop}%
\bibitem [{\citenamefont {Bordag}\ and\ \citenamefont
  {Khusnutdinov}(2008)}]{Bordag:2008:Ovesps}%
  \BibitemOpen
  \bibfield  {author} {\bibinfo {author} {\bibfnamefont {M.}~\bibnamefont
  {Bordag}}\ and\ \bibinfo {author} {\bibfnamefont {N.~R.}\ \bibnamefont
  {Khusnutdinov}},\ }\href {\doibase 10.1103/PhysRevD.77.085026} {\bibfield
  {journal} {\bibinfo  {journal} {Phys. Rev. D}\ }\textbf {\bibinfo {volume}
  {77}},\ \bibinfo {pages} {085026} (\bibinfo {year} {2008})}\BibitemShut
  {NoStop}%
\bibitem [{\citenamefont
  {Khusnutdinov}(2011)}]{Khusnutdinov:2011:vdWibaaaasps}%
  \BibitemOpen
  \bibfield  {author} {\bibinfo {author} {\bibfnamefont {N.~R.}\ \bibnamefont
  {Khusnutdinov}},\ }\href {\doibase 10.1103/PhysRevB.83.115454} {\bibfield
  {journal} {\bibinfo  {journal} {Phys. Rev. B}\ }\textbf {\bibinfo {volume}
  {83}},\ \bibinfo {pages} {115454} (\bibinfo {year} {2011})}\BibitemShut
  {NoStop}%
\bibitem [{\citenamefont {Khusnutdinov}(2012)}]{Khusnutdinov:2012:tCiasps}%
  \BibitemOpen
  \bibfield  {author} {\bibinfo {author} {\bibfnamefont {N.~R.}\ \bibnamefont
  {Khusnutdinov}},\ }\href {\doibase 10.1088/1751-8113/45/26/265301} {\bibfield
   {journal} {\bibinfo  {journal} {J. Phys. A: Math. Theor.}\ }\textbf
  {\bibinfo {volume} {45}},\ \bibinfo {pages} {265301} (\bibinfo {year}
  {2012})}\BibitemShut {NoStop}%
\bibitem [{\citenamefont {Abramowitz}\ and\ \citenamefont
  {Stegun}(1970)}]{Abramowitz:1970:eHMFFGMT}%
  \BibitemOpen
  \bibfield  {author} {\bibinfo {author} {\bibfnamefont {M.}~\bibnamefont
  {Abramowitz}}\ and\ \bibinfo {author} {\bibfnamefont {I.~A.}\ \bibnamefont
  {Stegun}},\ }\href@noop {} {\emph {\bibinfo {title} {Handbook of Mathematical
  Functions With Formulas, Graphs, and Mathematical Tables,}}},\ Applied
  Mathematics Series Vol. 55\ (\bibinfo  {publisher} {National Bureau of
  Standards, Washington, DC},\ \bibinfo {year} {1970})\BibitemShut {NoStop}%
\bibitem [{\citenamefont {Bateman}\ and\ \citenamefont
  {Erdelyi}(1953)}]{Bateman:1953:Htf-1}%
  \BibitemOpen
  \bibfield  {author} {\bibinfo {author} {\bibfnamefont {H.}~\bibnamefont
  {Bateman}}\ and\ \bibinfo {author} {\bibfnamefont {A.}~\bibnamefont
  {Erdelyi}},\ }\href@noop {} {\emph {\bibinfo {title} {Higher transcendental
  functions}}},\ Vol.~\bibinfo {volume} {1}\ (\bibinfo  {publisher}
  {McGraw-Hill, Inc., New York},\ \bibinfo {year} {1953})\ p.\ \bibinfo {pages}
  {296}\BibitemShut {NoStop}%
\bibitem [{\citenamefont {Boyer}(1968)}]{Boyer:1968:QEZEoaCSSatCMfaCP}%
  \BibitemOpen
  \bibfield  {author} {\bibinfo {author} {\bibfnamefont {T.~H.}\ \bibnamefont
  {Boyer}},\ }\href {\doibase 10.1103/PhysRev.174.1764} {\bibfield  {journal}
  {\bibinfo  {journal} {Phys. Rev.}\ }\textbf {\bibinfo {volume} {174}},\
  \bibinfo {pages} {1764} (\bibinfo {year} {1968})}\BibitemShut {NoStop}%
\bibitem [{\citenamefont {Vassilevich}(2003)}]{Vassilevich:2003:HkeUm}%
  \BibitemOpen
  \bibfield  {author} {\bibinfo {author} {\bibfnamefont {D.~V.}\ \bibnamefont
  {Vassilevich}},\ }\href {\doibase 10.1016/j.physrep.2003.09.002} {\bibfield
  {journal} {\bibinfo  {journal} {Phys. Rept.}\ }\textbf {\bibinfo {volume}
  {388}},\ \bibinfo {pages} {279} (\bibinfo {year} {2003})}\BibitemShut
  {NoStop}%
\bibitem [{\citenamefont {Bulgac}\ \emph
  {et~al.}(2006{\natexlab{b}})\citenamefont {Bulgac}, \citenamefont
  {Magierski},\ and\ \citenamefont {Wirzba}}]{Bulgac:2006:SCebDsoapaas}%
  \BibitemOpen
  \bibfield  {author} {\bibinfo {author} {\bibfnamefont {A.}~\bibnamefont
  {Bulgac}}, \bibinfo {author} {\bibfnamefont {P.}~\bibnamefont {Magierski}}, \
  and\ \bibinfo {author} {\bibfnamefont {A.}~\bibnamefont {Wirzba}},\ }\href
  {\doibase 10.1103/PhysRevD.73.025007} {\bibfield  {journal} {\bibinfo
  {journal} {Phys. Rev. D}\ }\textbf {\bibinfo {volume} {73}},\ \bibinfo
  {pages} {025007} (\bibinfo {year} {2006}{\natexlab{b}})}\BibitemShut
  {NoStop}%
\end{thebibliography}
